\pdfoutput=1
\documentclass[%
 reprint,showpacs,
 superscriptaddress,
 amsmath,amssymb,
 aps,
 prl,
]{revtex4-1}
\usepackage{longtable}
\usepackage{graphicx}
\usepackage{dcolumn}
\usepackage{bm}
\usepackage[
   bookmarks=true,
   colorlinks=true,
   hyperindex=true,
   citecolor=blue,
   linkcolor=blue,
   urlcolor=blue
]{hyperref}
\usepackage{float}
\usepackage{titlesec}

\begin{document}
\title{ZAIGA: Zhaoshan Long-baseline Atom Interferometer Gravitation Antenna}

\author{Ming-Sheng Zhan}
\email{mszhan@wipm.ac.cn}
\affiliation{State Key Laboratory of Magnetic Resonance and Atomic and Molecular Physics, Wuhan Institute of Physics and Mathematics, Chinese Academy of Sciences, Wuhan 430071, China}
\affiliation{Center for Cold Atom Physics, Chinese Academy of Sciences, Wuhan 430071, China}
\affiliation{School of Physics, University of Chinese Academy of Sciences, Beijing 100049, China}
\author{Jin Wang}
\email{wangjin@wipm.ac.cn}
\affiliation{State Key Laboratory of Magnetic Resonance and Atomic and Molecular Physics, Wuhan Institute of Physics and Mathematics, Chinese Academy of Sciences, Wuhan 430071, China}
\affiliation{Center for Cold Atom Physics, Chinese Academy of Sciences, Wuhan 430071, China}
\affiliation{School of Physics, University of Chinese Academy of Sciences, Beijing 100049, China}
\author{Wei-Tou Ni}
\affiliation{State Key Laboratory of Magnetic Resonance and Atomic and Molecular Physics, Wuhan Institute of Physics and Mathematics, Chinese Academy of Sciences, Wuhan 430071, China}
\author{Dong-Feng Gao}
\affiliation{State Key Laboratory of Magnetic Resonance and Atomic and Molecular Physics, Wuhan Institute of Physics and Mathematics, Chinese Academy of Sciences, Wuhan 430071, China}
\affiliation{Center for Cold Atom Physics, Chinese Academy of Sciences, Wuhan 430071, China}
\affiliation{School of Physics, University of Chinese Academy of Sciences, Beijing 100049, China}
\author{Gang Wang}
\affiliation{State Key Laboratory of Magnetic Resonance and Atomic and Molecular Physics, Wuhan Institute of Physics and Mathematics, Chinese Academy of Sciences, Wuhan 430071, China}
\author{Ling-Xiang He}
\affiliation{State Key Laboratory of Magnetic Resonance and Atomic and Molecular Physics, Wuhan Institute of Physics and Mathematics, Chinese Academy of Sciences, Wuhan 430071, China}
\affiliation{Center for Cold Atom Physics, Chinese Academy of Sciences, Wuhan 430071, China}
\affiliation{School of Physics, University of Chinese Academy of Sciences, Beijing 100049, China}
\author{Run-Bing Li}
\affiliation{State Key Laboratory of Magnetic Resonance and Atomic and Molecular Physics, Wuhan Institute of Physics and Mathematics, Chinese Academy of Sciences, Wuhan 430071, China}
\affiliation{Center for Cold Atom Physics, Chinese Academy of Sciences, Wuhan 430071, China}
\affiliation{School of Physics, University of Chinese Academy of Sciences, Beijing 100049, China}
\author{Lin Zhou}
\affiliation{State Key Laboratory of Magnetic Resonance and Atomic and Molecular Physics, Wuhan Institute of Physics and Mathematics, Chinese Academy of Sciences, Wuhan 430071, China}
\affiliation{Center for Cold Atom Physics, Chinese Academy of Sciences, Wuhan 430071, China}
\affiliation{School of Physics, University of Chinese Academy of Sciences, Beijing 100049, China}
\author{Xi Chen}
\affiliation{State Key Laboratory of Magnetic Resonance and Atomic and Molecular Physics, Wuhan Institute of Physics and Mathematics, Chinese Academy of Sciences, Wuhan 430071, China}
\affiliation{Center for Cold Atom Physics, Chinese Academy of Sciences, Wuhan 430071, China}
\affiliation{School of Physics, University of Chinese Academy of Sciences, Beijing 100049, China}
\author{Jia-Qi Zhong}
\affiliation{State Key Laboratory of Magnetic Resonance and Atomic and Molecular Physics, Wuhan Institute of Physics and Mathematics, Chinese Academy of Sciences, Wuhan 430071, China}
\affiliation{Center for Cold Atom Physics, Chinese Academy of Sciences, Wuhan 430071, China}
\affiliation{School of Physics, University of Chinese Academy of Sciences, Beijing 100049, China}
\author{Biao Tang}
\affiliation{State Key Laboratory of Magnetic Resonance and Atomic and Molecular Physics, Wuhan Institute of Physics and Mathematics, Chinese Academy of Sciences, Wuhan 430071, China}
\affiliation{Center for Cold Atom Physics, Chinese Academy of Sciences, Wuhan 430071, China}
\affiliation{School of Physics, University of Chinese Academy of Sciences, Beijing 100049, China}
\author{Zhan-Wei Yao}
\affiliation{State Key Laboratory of Magnetic Resonance and Atomic and Molecular Physics, Wuhan Institute of Physics and Mathematics, Chinese Academy of Sciences, Wuhan 430071, China}
\affiliation{Center for Cold Atom Physics, Chinese Academy of Sciences, Wuhan 430071, China}
\affiliation{School of Physics, University of Chinese Academy of Sciences, Beijing 100049, China}
\author{Lei Zhu}
\affiliation{State Key Laboratory of Magnetic Resonance and Atomic and Molecular Physics, Wuhan Institute of Physics and Mathematics, Chinese Academy of Sciences, Wuhan 430071, China}
\affiliation{Center for Cold Atom Physics, Chinese Academy of Sciences, Wuhan 430071, China}
\author{Zong-Yuan Xiong}
\affiliation{State Key Laboratory of Magnetic Resonance and Atomic and Molecular Physics, Wuhan Institute of Physics and Mathematics, Chinese Academy of Sciences, Wuhan 430071, China}
\affiliation{Center for Cold Atom Physics, Chinese Academy of Sciences, Wuhan 430071, China}
\author{Si-Bin Lu}
\affiliation{State Key Laboratory of Magnetic Resonance and Atomic and Molecular Physics, Wuhan Institute of Physics and Mathematics, Chinese Academy of Sciences, Wuhan 430071, China}
\affiliation{Center for Cold Atom Physics, Chinese Academy of Sciences, Wuhan 430071, China}
\author{Geng-Hua Yu}
\affiliation{State Key Laboratory of Magnetic Resonance and Atomic and Molecular Physics, Wuhan Institute of Physics and Mathematics, Chinese Academy of Sciences, Wuhan 430071, China}
\affiliation{Center for Cold Atom Physics, Chinese Academy of Sciences, Wuhan 430071, China}
\author{Qun-Feng Cheng}
\affiliation{State Key Laboratory of Magnetic Resonance and Atomic and Molecular Physics, Wuhan Institute of Physics and Mathematics, Chinese Academy of Sciences, Wuhan 430071, China}
\affiliation{Center for Cold Atom Physics, Chinese Academy of Sciences, Wuhan 430071, China}
\affiliation{School of Physics, University of Chinese Academy of Sciences, Beijing 100049, China}
\author{Min Liu}
\affiliation{State Key Laboratory of Magnetic Resonance and Atomic and Molecular Physics, Wuhan Institute of Physics and Mathematics, Chinese Academy of Sciences, Wuhan 430071, China}
\affiliation{Center for Cold Atom Physics, Chinese Academy of Sciences, Wuhan 430071, China}
\affiliation{School of Physics, University of Chinese Academy of Sciences, Beijing 100049, China}
\author{Yu-Rong Liang}
\affiliation{State Key Laboratory of Magnetic Resonance and Atomic and Molecular Physics, Wuhan Institute of Physics and Mathematics, Chinese Academy of Sciences, Wuhan 430071, China}
\affiliation{Center for Cold Atom Physics, Chinese Academy of Sciences, Wuhan 430071, China}
\affiliation{School of Physics, University of Chinese Academy of Sciences, Beijing 100049, China}
\author{Peng Xu}
\affiliation{State Key Laboratory of Magnetic Resonance and Atomic and Molecular Physics, Wuhan Institute of Physics and Mathematics, Chinese Academy of Sciences, Wuhan 430071, China}
\affiliation{Center for Cold Atom Physics, Chinese Academy of Sciences, Wuhan 430071, China}
\affiliation{School of Physics, University of Chinese Academy of Sciences, Beijing 100049, China}
\author{Xiao-Dong He}
\affiliation{State Key Laboratory of Magnetic Resonance and Atomic and Molecular Physics, Wuhan Institute of Physics and Mathematics, Chinese Academy of Sciences, Wuhan 430071, China}
\affiliation{Center for Cold Atom Physics, Chinese Academy of Sciences, Wuhan 430071, China}
\affiliation{School of Physics, University of Chinese Academy of Sciences, Beijing 100049, China}
\author{Min Ke}
\affiliation{State Key Laboratory of Magnetic Resonance and Atomic and Molecular Physics, Wuhan Institute of Physics and Mathematics, Chinese Academy of Sciences, Wuhan 430071, China}
\affiliation{Center for Cold Atom Physics, Chinese Academy of Sciences, Wuhan 430071, China}
\affiliation{School of Physics, University of Chinese Academy of Sciences, Beijing 100049, China}
\author{Zheng Tan}
\affiliation{State Key Laboratory of Magnetic Resonance and Atomic and Molecular Physics, Wuhan Institute of Physics and Mathematics, Chinese Academy of Sciences, Wuhan 430071, China}
\affiliation{Center for Cold Atom Physics, Chinese Academy of Sciences, Wuhan 430071, China}
\author{Jun Luo}
\affiliation{State Key Laboratory of Magnetic Resonance and Atomic and Molecular Physics, Wuhan Institute of Physics and Mathematics, Chinese Academy of Sciences, Wuhan 430071, China}
\affiliation{Center for Cold Atom Physics, Chinese Academy of Sciences, Wuhan 430071, China}
\affiliation{School of Physics, University of Chinese Academy of Sciences, Beijing 100049, China}

\date{\today}

\begin{abstract}
The Zhaoshan long-baseline Atom Interferometer Gravitation Antenna (ZAIGA) is a new type of underground laser-linked interferometer facility, and is currently under construction. It is in the 200-meter-on-average underground of a mountain named Zhaoshan which is about 80 km southeast to Wuhan. ZAIGA will be equipped with long-baseline atom interferometers, high-precision atom clocks, and large-scale gyros. ZAIGA facility will take an equilateral triangle configuration with two 1-km-apart atom interferometers in each arm, a 300-meter vertical tunnel with atom fountain and atom clocks mounted, and a tracking-and-ranging 1-km-arm-length prototype with lattice optical clocks linked by locked lasers. The ZAIGA facility will be used for experimental research on gravitation and related problems including gravitational wave detection, high-precision test of the equivalence principle of micro-particles, clock based gravitational red-shift measurement, rotation measurement and gravito-magnetic effect.

\end{abstract}



\maketitle

\section{1. Introduction}	

The unification of fundamental interactions is the major frontier of sciences to be broken. Although, on one hand, quantum theory has achieved great success in describing microscopic physical phenomena; on the other hand, general relativity (GR) has withstood long-term tests in describing the gravitational interactions of macroscopic objects, and many theoretical frameworks have been proposed to try to unify quantum theory and GR. However, due to the lack of experimental evidence to these theories, a theory of everything still needs to be established. At present time, in the experimentally controllable parameter space, the gravity and other three (strong-, weak-, electromagnetic-) interactions are difficult to intersect (coupling), just like trains running in two parallel orbits. Research and progress in the areas where intersections may occur, such as the gravitational effects of microscopic systems or the quantum effects of macroscopic objects, have received much attention.


The in-depth study of gravitation and related theories has important scientific significance for understanding the law of gravity. For example, the equivalence principle is the basis of Einstein's GR, the equivalence principle test is a direct test of basic assumptions of GR; The gravitational wave detection can not only deeply verify Einstein's GR, but also have great potential for acquiring information about the Universe. The study of these fundamental physics plays an important role in the development of science and technology and the progress of human society.

At present, scientists continue to test the existing physical laws and explore their applicable range from two extremes, high energy (high energy physics) and large scale (cosmology). On the other hand, with the development of quantum control methods and techniques at atomic and molecular scale, the measurement of the microscopic system has reached an unprecedented high precision, so it is possible to find fine traces of new physics in low-energy atomic systems through precision measurement.

Long-baseline interferometer undoubtedly represents the precision and difficulty of precision measurements. From the radio Very Long Baseline Interferometer (VLBI) for astronomical observation to the Laser Interference Gravitational Wave Observatory (LIGO) for gravitational wave detection, all these facilities have experienced years of technical accumulation and development, and have greatly promoted the scientific progress. The atom interferometer (AI) is a new instrument for precision measurement, that can be used to measure physical parameters such as rotation, acceleration, electric field, magnetic field, and gravity. Since 1991, AIs have been gradually applied to geodesy, inertial navigation, general relativity testing, fundamental physics constant determination, and so on.

The measurement accuracy of an AI is limited by many technical parameters, and the free fall time of atoms is an important factor among them. The free evolution time of ground-based AIs is currently on the level of 10 - 1000 ms. The space microgravity environment can prolong the free evolution of atoms\cite{LiuL2018}. To build a 100-meter-level long-baseline AI on the ground is suitable to carry out precision measurement and gravitation research, where the free evolution time of atoms is comparable to that in space, which breaks through the environment limitations of the space station.

The Zhaoshan long-baseline Atom Interferometer Gravitation Antenna (ZAIGA) is a new type of underground laser-linked interferometer facility, and is currently under construction. It is in the 200-meter-on-average underground of a mountain named Zhaoshan which is about 80 km southeast to Wuhan. ZAIGA will be based on matter wave interference, which is different from the existing long baseline laser interferometers in terms of principle, technology and application. ZAIGA will be equipped with the latest atom interferometry and will take an equilateral triangle configuration, with two 1-km-apart atom interferometers in each arm.
\begin{figure}[htbp]
\includegraphics[width=8cm]{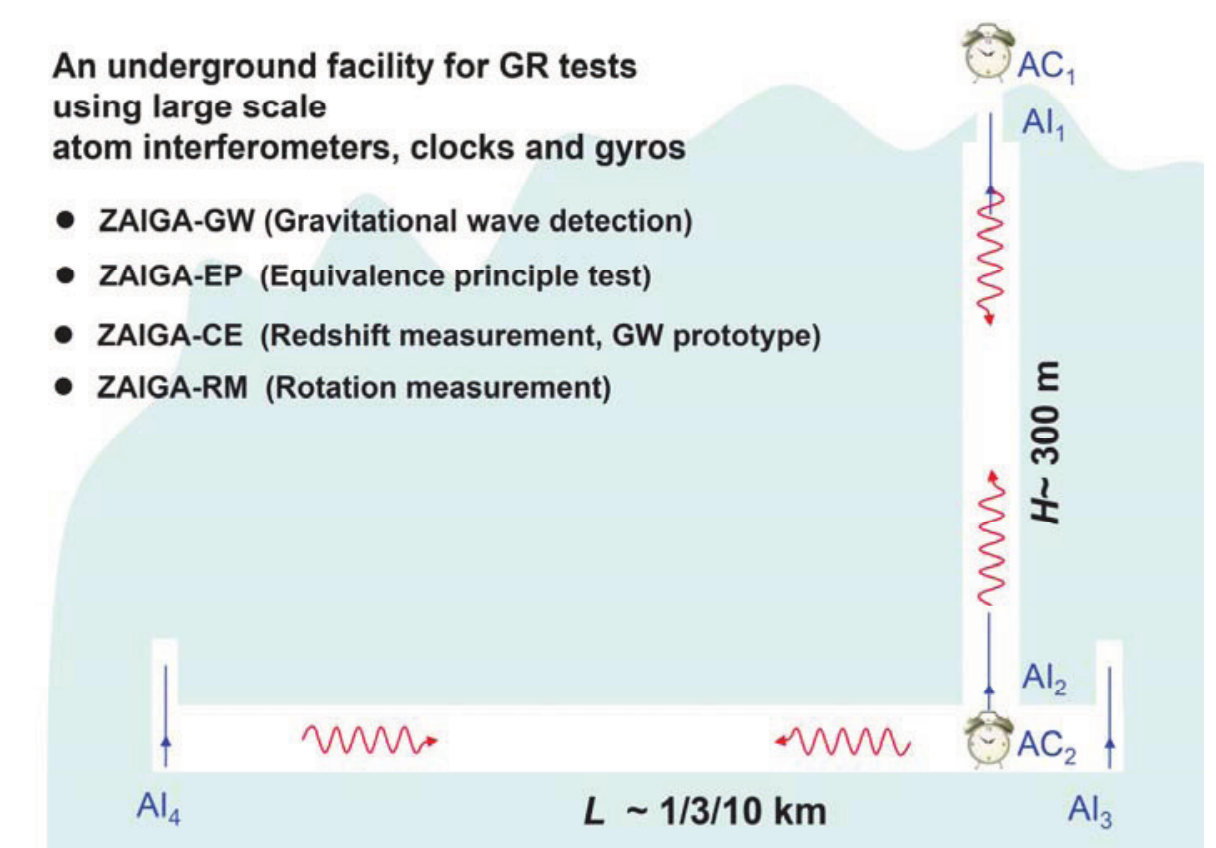}
\caption{Schematic diagram of ZAIGA.}
\label{f1.2}
\end{figure}

\begin{figure}[htbp]
\centering
\includegraphics[width=8cm]{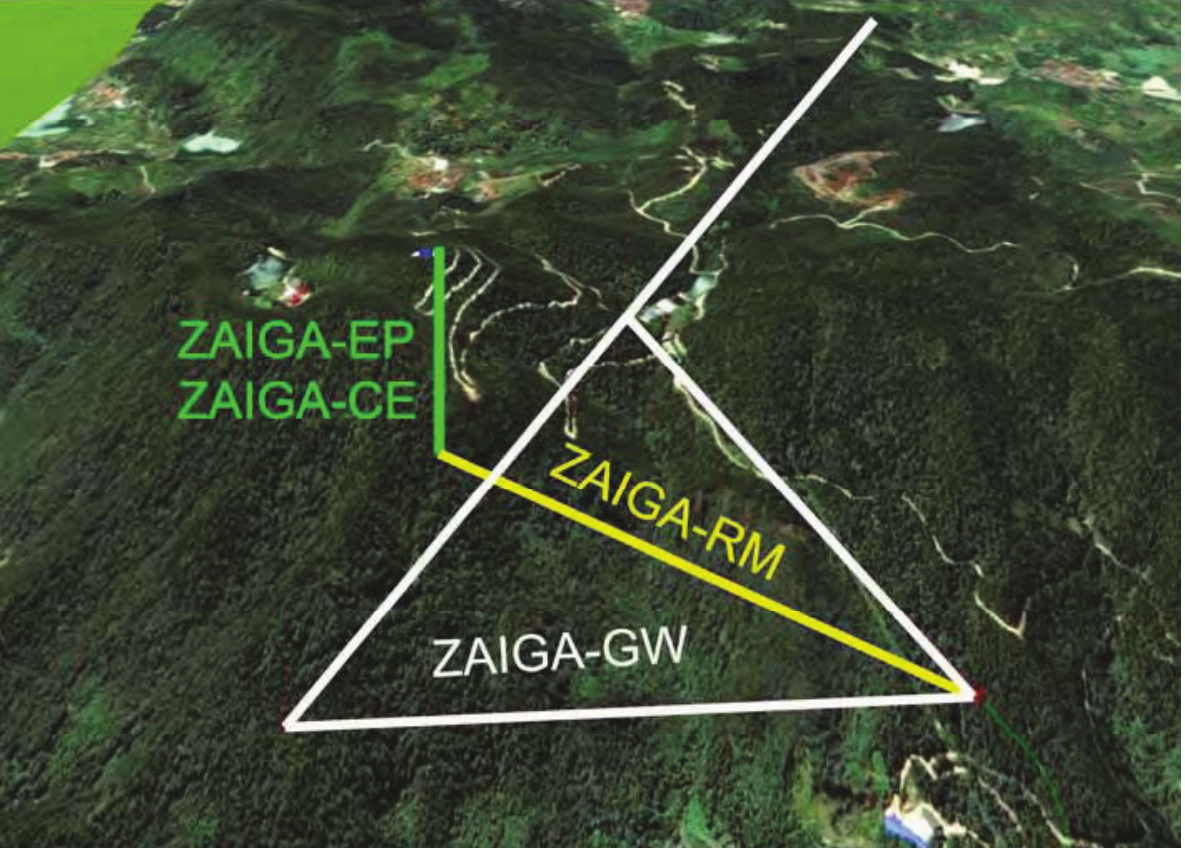}
\caption{The bird's eye view of the ZAIGA facility.}
\label{f1.3}
\end{figure}

\begin{figure}[htbp]
\includegraphics[width=8cm]{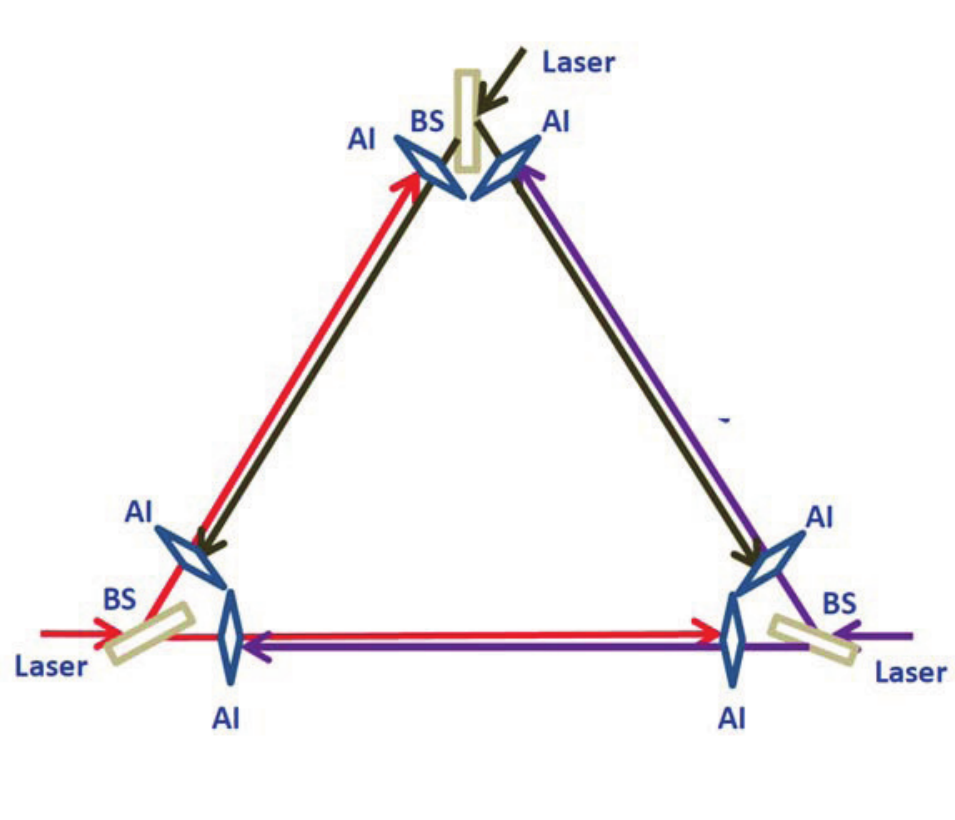}
\caption{The conceptual diagram of the ZAIGA-GW.}
\label{f1.4}
\end{figure}

The ZAIGA facility will be used for experimental research on gravitation and related problems. The schematic diagram of ZAIGA is depicted in Fig.~\ref{f1.2}, it includes five systems: ZAIGA-GW for gravitational wave detection, ZAIGA-EP for high-precision test of the equivalence principle of microparticles, ZAIGA-CE for clock based gravitational redshift measurement and gravitational wave detection, ZAIGA-CE-GW for tracking-and-ranging 1-km-arm-length prototype with lattice optical clocks linked by locked lasers, ZAIGA-RM for rotation gravitomagnetism measurement. The bird's eye view of the ZAGA facility is shown in Fig.~\ref{f1.3}, and the conceptual diagram of the ZAIGA-GW is shown in Fig.~\ref{f1.4}.

\section{2. ZAIGA-GW: Gravitational Wave Detection}

According to Einstein's theory of general relativity, the disturbance in the spacetime curvature will propagate in the form of wave, which is called the gravitational wave (GW). The first indirect evidence of its existence comes from the timing observations of the binary pulsar, PSR 1913+16 \cite{taylor2010}. Since 1960s, a lot of efforts on direct GW detection have been made, and many ground-based GW detectors have been built \cite{ju2000}. After decades of improvement in sensitivity, the LIGO made the first two direct detections of high frequency GWs in 2015 \cite{ligo2016a,ligo2016b}, and three more GW detections were announced in 2017 \cite{ligo2017a,ligo2017b, ligo2017c}. Then, a new era in the study of the universe started. However, since the frequency band of the ground-based laser interferometer detectors is bounded by the seismic and Newtonian noise from below, it is difficult to detect GWs with frequencies lower than a few Hz, even for the so-called third-generation GW detectors---the Einstein Telescope \cite{et2011a} and the Cosmic Explorer \cite{ce2017}. Physically, it is of great importance to extend the GW detection into the lower frequency bands \cite{nibook}. To do so, one way is to avoid the Earth's environment, and go into space, the other way is to develop methods of dealing with the Earth's environmental noise, and build terrestrial GW detectors based on technologies other than laser interferometry. Several space-based laser interferometer GW detection schemes (such as LISA \cite{lisa2011}, Taiji \cite{taiji2018}, TianQin \cite{tianqin2016}, DECIGO \cite{decigo2010}, BBO \cite{bbo2003}, and AMIGO \cite{ni2017}) have been put forward. But, to reach necessary sensitivity with wide frequency range, these space-based laser interferometer schemes typically have to be large in size, expensive to build, and relatively short in observation.

Actually, after more than twenty years of development, the rapid advances in atom interferometry provide us such a possibility. Atom interferometers (AIs) have already reached a very high level of sensitivity \cite{cronin2009}, and have been used in a wide variety of precision measurements. Many important experimental results have been made. For example, the measurements of the fine structure constant in atom-recoil experiments are the most accurate ones to date \cite{clade2011,muller2018}. The measurement of the Newtonian gravitational constant, with a double atom-interferometer-gravity-gradiometer \cite{tino2014}, was adopted as one of the 14 measured values in the 2014 CODATA \cite{codata2014}. The test of the weak equivalence principle (WEP) in a double-species-AI experiment reached the $10^{-8}$ level\cite{zhan2015}.

Inspired by these impressive developments, people began to investigate the possibility of detecting GWs with AIs. In recent years, many experimental GW detection schemes have been proposed and discussed \cite{chiao2004,roura2006,tino2007,kasevich2008,hogen2011,gao2011,harms2013,kasevich2016,chaibi2016,gao2018}. Depending on the role of AIs in the GW detection, these schemes can be classified into two types. The first type is the GW detection with a single AI \cite{chiao2004,roura2006,tino2007,gao2011,gao2018}. It makes use of the strain effect on the atom's trajectory induced by GWs. This effect will generate a tiny phase shift, which is proportional to the size of the AI. To enhance this tiny effect, the size of the AI has to be large. However, due to technical limitations, the size of an AI could not be made as large as one wishes. The second type is the GW detection based on laser-linked AIs \cite{kasevich2008,hogen2011,harms2013,kasevich2016,chaibi2016}. It mainly focuses on the GW-induced strain effect on the trajectory of lights. This strain effect is then transferred to the AIs through interactions between lights and atoms. The resulting phase shift is proportional to the distance between two laser-linked AIs. Technically, it is easier to increase the distance between AIs than the size of an AI.

The ZAIGA-GW is a proposed underground GW detector in an equilateral triangle configuration. To minimize the effect of the seismic noise, the whole detector is about 200-meter-on-average below a mountain. With two 3-km-apart laser-linked AIs, each arm can detect the GW independently. In other words, the ZAIGA-GW actually consists of three co-located GW detectors in a closed triangle. Thus, the ZAIGA-GW has several similar advantages as the Einstein Telescope \cite{ET2009}. The first advantage is the generation of null data stream, which is very important for the identification of noise and false GW candidates. The second advantage is that three co-located GW detectors in a closed triangle provide a full detection of both GW polarizations, $h_{+}$ and $h_{\times}$. The last advantage is on the operational aspect. Since only two out of the three detectors are independent, the maintenance of one detector does not interrupt the observation. Therefore, in principle, the ZAIGA-GW could have a $100\%$ duty cycle.

\subsection{2.1 Overall Description of ZAIGA-GW}

The schematic diagram of ZAIGA-GW is depicted in Fig.~\ref{f1.3}. At the initial stage, only one side and a small triangle of 1 km side length will be built, which is indicated by the white solid lines in Fig.~\ref{f1.3}.  ZAIGA-GW could be upgraded to 3 km arms in future, and even to 10 km in far future. As shown in Fig.~\ref{f1.4}, at each vertex of the triangle, the laser light beam is split into two beams by the splitter. Thus, in each side, there are two counter-propagating light beams, which are used to control the two AIs.

\begin{figure}[htpb]
\includegraphics[width=8.0cm]{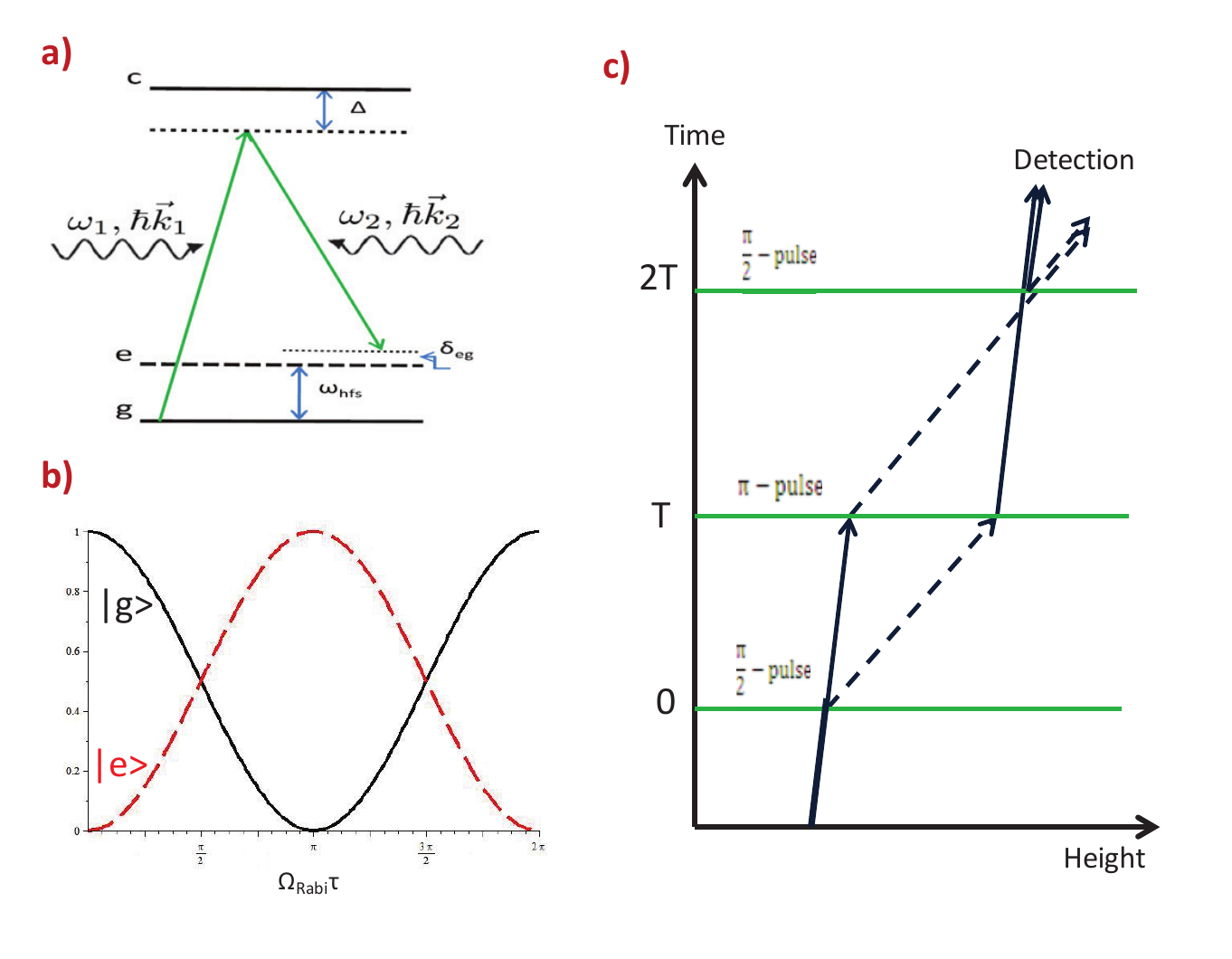}
\caption{(a) The diagram for a stimulated Raman transition between two atomic hyperfine ground states $|g\rangle$ and $ |e\rangle$. Laser beams are detuned by $\Delta$ from the optical state $|c\rangle$. The atomic population is resonantly transferred between $|g\rangle$ and $|e\rangle$ if the frequency difference $\omega_1-\omega_2$ is close to $\omega_{{\rm hfs}}$. (b) A $\frac{\pi}{2}$-pulse is a beam splitter since the atom beam prepared in one state is transferred into a superposition of states $ |g\rangle$  and $ |e\rangle$. A $\pi$-pulse is a beam reflecter since the atomic state is reversed completely. (c) The light pulse sequence for a $\frac{\pi}{2}$-$\pi$-$\frac{\pi}{2}$ Mach-Zender atom interferometer. \label{f2.1}}
\end{figure}

A typical ${\pi}$/2-$\pi$-${\pi}$/2 light pulse atom interferometer is depicted in Fig.~\ref{f2.1}, and its theory can be found in early paper\cite{chu1992}.  The stimulated Raman transitions are realized by two counter-propagating laser beams. One light beam from the control laser, with frequency $\omega_1$ and wave vector ${\bf \emph{k}_1}$, is always on. The other light beam from the passive laser, with frequency $\omega_2$ and wave vector ${\bf \emph{k}_2}$, is triggered by a timer. First, cold atom clouds with as many atoms as possible, prepared in the $|g\rangle$ state by a two-dimensional magneto-optical trap (2D-MOT), are loaded into the three-dimensional magneto-optical trap (3D-MOT). At time $t=0$, the first Raman $\pi$/2 pulse is applied to the atomic beam, and coherently splits the atomic wave packet into a superposition of states $|g\rangle$ and $|e\rangle$, which differ in momentum by $\hbar {\bf \emph{k}}_{{\rm eff}}$, where ${\bf \emph{k}}_{{\rm eff}}={\bf \emph{k}_2}-{\bf \emph{k}_1}\simeq 2 {\bf \emph{k}_2}$. Atoms travel along the free-fall paths, and the beams in different atomic states will spatially separate due to difference in momentum. After a evolution time $T$, the Raman $\pi$ pulses are applied, which transit the state $|g\rangle$ to $|e\rangle$ and the state $|e\rangle$ to $|g\rangle$, respectively. Then, after another evolution time $T$, the two wave packets overlap, and the last Raman ${\pi}$/2 pulses are applied to complete the interference. Finally, the signal can be measured by detecting the number of atoms in either $|g\rangle$ or $|e\rangle$ states.

Take the triangle to be in the $xy$ plane, and suppose the GWs propagate in the $z$ direction. In the transverse traceless (TT) gauge, the metric for the GWs is written as
\begin{equation}
\begin{split}
ds^2 & =(\eta_{\mu\nu}+h_{\mu\nu})\, dx^{\mu}dx^{\nu} \\
& =-dt^2+ (\delta_{ij}+h_{ij}) \, dx^idx^j,
\end{split}
\end{equation}
where $z\equiv x^3$, $h^i_i=0$, $\partial^ih_{ij}=0$ and $i, j=1,2,3$. If the only none-zero components of the metric are $h_{11}=-h_{22}=h \, {\rm e}^{i (2\pi f t + \phi_0)}$, then it is called the $h_+$-polarization. Accordingly, the metric with none-zero $h_{12}=h_{21}=h \, {\rm e}^{i (2\pi f t + \phi_0)}$ is called the $h_\times$-polarization. $h$ and $f$ are the amplitude and frequency of the GW, respectively. $\phi_0$ is the initial phase of the GW at the time of arriving at the AIs.

The atom's phase in an AI mainly contains the propagation phase along the atomic paths, and the transferred phase through interaction with laser pulses.\cite{kasevich2008} Since it takes $L/c$ time for the laser light to interact with the first AI and the second AI, the laser phases transferred to the two AIs will roughly differ by $k_2 L$ at each interaction. When the GWs go into the detector, the arm length will change in time, which will result in the change in the laser phase transferred to the two AIs. With a careful and lengthy calculation, when $f \ll c/L$, the differential phase shift between the two AIs is simplified into
\begin{equation}
\delta \phi_{{\rm AI}}= 2 k_{{\rm eff}} \, h\, L \,  {\rm sin}^2\left(\pi f T\right) {\rm sin}(\phi_0).
\label{phaseshift}
\end{equation}
One can see that the detector has the best sensitivity at frequencies $f=\frac{2n-1}{2T}$, loses sensitivity at frequencies $f=\frac{n}{T}$. These frequencies are determined by $T$, i.e., the height of AIs. Typically, $h < 10^{-21}$, the phase shift in Eq. (\ref{phaseshift}) is very small. To enhance it, one can either enlarge $L$, or use the so-called large momentum transfer (LMT) technology to increase $k_{{\rm eff}}$\cite{chiow2011}.

\subsection{2.2 The sensitivity and technical requirements of ZAIGA-GW}

The detection sensitivity of ZAIGA-GW is determined by various noise sources, where different noise dominates in different frequency range. These noises come from either the detector itself, or the environment where the detector is installed. In the 0.1$\sim$10 Hz frequency range, as discussed in previous literatures \cite{kasevich2008,harms2013,chaibi2016,Canuel2018}, the detection sensitivity is mainly limited by the shot noise and the Newtonian noise (NN).

\subsubsection{2.2.1 Atomic shot noise}
One fundamental noise from the detector is the shot noise of AIs. For ZAIGA-GW, the power spectral density (PSD) for the shot noise, $\tilde{h}_{sh} (f)$, is easily derived from Eq. (\ref{phaseshift}),
\begin{eqnarray}
 \tilde{h}_{sh} (f)=\frac{1}{2 k_{{\rm eff}} \, L \,  {\rm sin}^2\left(\pi f T\right)\sqrt{\mathcal{R}}},
\end{eqnarray}
where $\mathcal{R}$ is the flux intensity of the atom beam.

\begin{figure}[htpb]
\includegraphics[width=8.0cm]{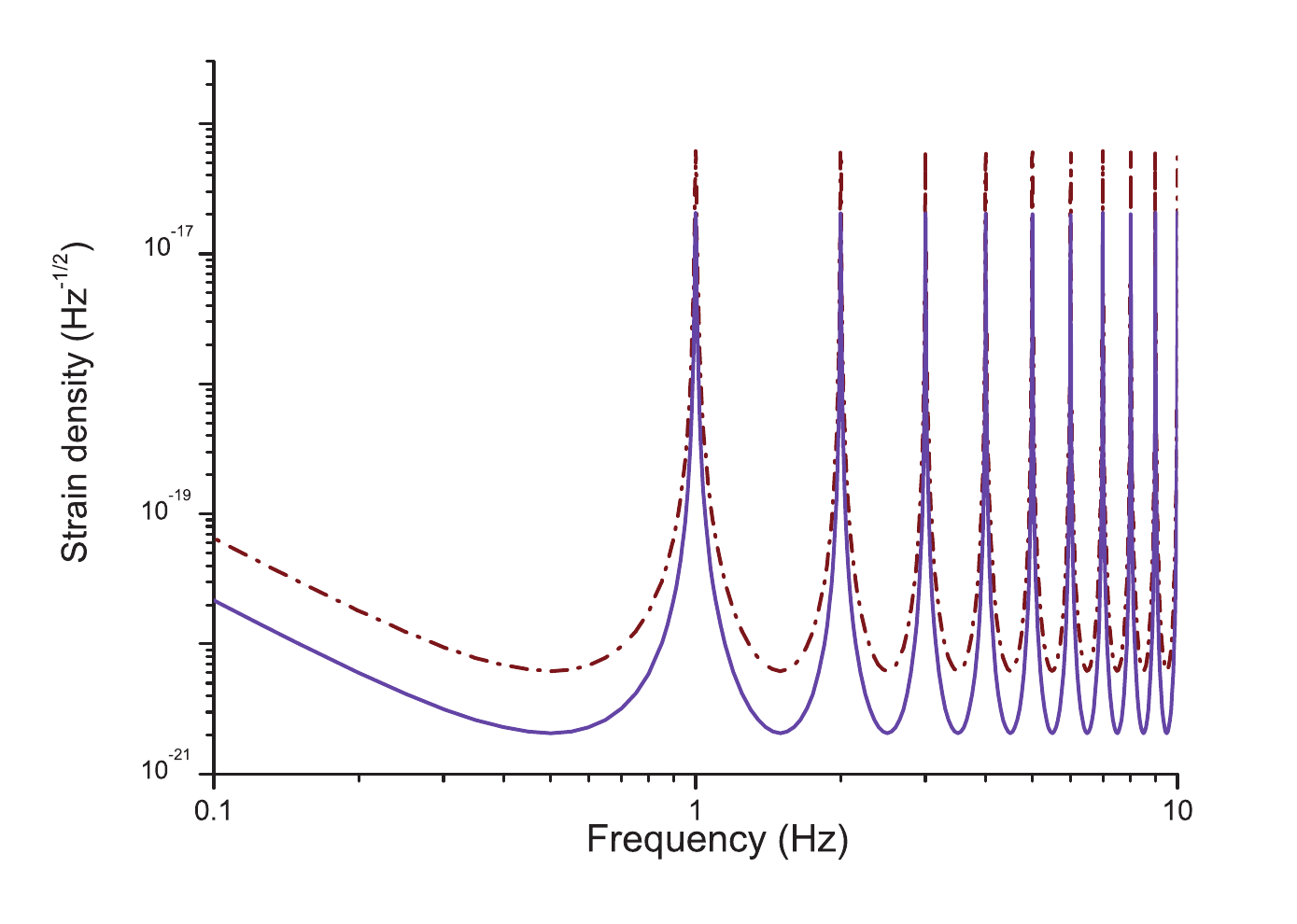}
\caption{Power spectral density curves for shot noise of ZAIGA-GW. The dash-dot line is for $L$ = 1 km, and the solid line is for $L$ = 3 km. \label{f2.2}}
\end{figure}

Take the 5-meter-high (i.e., $T$ = 1 s) $^{87}$Rb AI as an example, where the 5$S_{1/2} |F=2, m_F = 0\rangle $ and 5$S_{1/2} |F=1, m_F = 0\rangle $ states are adopted as $ |g\rangle$ and $ |e\rangle$ states, respectively. To split, reflect, and recombine the $^{87}$Rb atom beams, the Raman pulses with the wavelength $\lambda_l = 780$ nm are used. Assuming that the $N$ = 1000 LMT beam splitter ($k_{{\rm eff}}=2Nk_l=1.6\times 10^{10}\, {\rm m^{-1}}$) and the flux intensity $\mathcal{R}=10^{14}$ atoms/s are realizable, we draw the corresponding sensitivity curves in Fig.~\ref{f2.2}. The sensitivity curves are truncated at 10 Hz, which is the data taking rate $f_{\rm d}$.  As discussed by Dimopoulos {\it et al}. \cite{kasevich2008}, $f_{\rm d}$ is the frequency of running the atom beams through the interferometer, which is no higher than 1 Hz.

\subsubsection{2.2.2 Newtonian noise}

\begin{figure}[htpb]
\includegraphics[width=4.0cm]{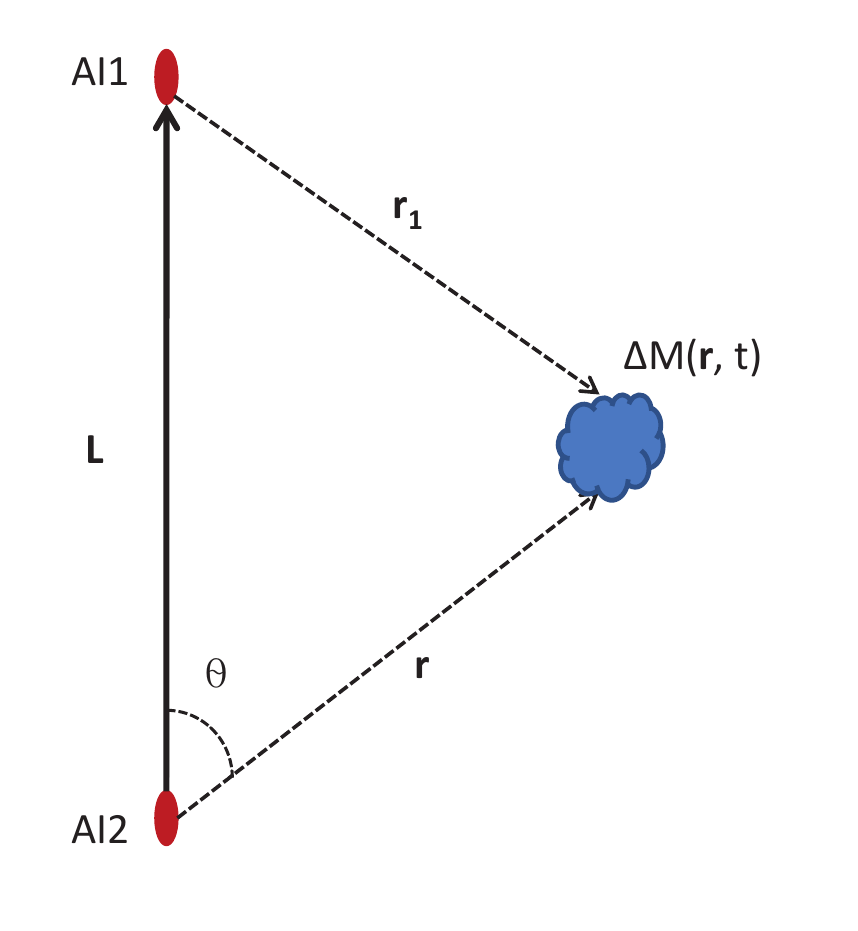}
\caption{The Newtonian noise to the detector caused by mass fluctuation in a nearby volume, where the two AIs are separated by a distance $L$. \label{f2.3}}
\end{figure}

For all types of ground-based GW detectors, the NN, also called the gravity gradient noise, is an unavoidable problem on the detection sensitivity below 10 Hz. It could not be eliminated by isolating the test masses from the environment, because the NN is caused by the direct gravitational coupling of the test masses with the nearby fluctuating terrestrial gravity field. The sources for the NN are mainly from the ambient seismic fields and the atmospheric pressure fluctuations, which are called seismic Newtonian noise (SNN) and infrasound Newtonian noise (INN), respectively. Following the pioneering work \cite{saulson1984}, people have done a lot of work on the NN in GW detectors \cite{hughes1998, beker2012, vetrano2013, harms2015}. Using the existing methods \cite{saulson1984, vetrano2013,Junca2019}, we can estimate the NN of ZAIGA-GW.
As illustrated in Fig.~\ref{f2.3}, a mass fluctuation $\Delta M({\bf r}, t)$ inside a volume element at point ${\bf r}$ will cause fluctuations in the acceleration field at positions AI$_{1}$ and AI$_{2}$ as
\begin{eqnarray*}
{\bf g}_1({\bf r}, t) &=& \frac{G \Delta M({\bf r}, t)}{{\bf r}_1^3}{\bf r}_1=\frac{G \Delta M({\bf r}, t)}{|{\bf r}-{\bf L}|^3}({\bf r}-{\bf L}), \\
{\bf g}_2({\bf r}, t) &=& \frac{G \Delta M({\bf r}, t)}{{\bf r}^3}{\bf r} \, ,
\end{eqnarray*}
where $G$ is the gravitational constant.

The resulting NN along the ${\bf \emph{L}}$ direction in the frequency domain is
\begin{equation}
\tilde{h}_{NN}(\omega , {\bf r})= \frac{2}{\omega^2 L}\left[ \tilde{g}_2(\omega , {\bf r})-\tilde{g}_1(\omega , {\bf r}) \right].
\end{equation}

To obtain the total NN, $\tilde{h}_{NN}(\omega)$, we need to sum over the whole space around the two AIs. Suppose that the space is homogeneous, and that the mass fluctuations do not depend on $\emph{r}$. We have
\begin{widetext}
\begin{equation}
\tilde{h}^2_{NN}(\omega)= \frac{4G^2\Delta\tilde{M}^2(\omega)}{\omega^4 L^2}\left(\frac{2}{\lambda_c}\right)^3 \int \left[ \frac{{\rm cos}\theta}{r^2}- \frac{r {\rm cos}\theta-L}{(r^2+L^2-2 r L {\rm cos}\theta)^{3/2}} \right]^2 r^2 dr\, d{\rm cos}\theta \, d\phi \, .
\label{eqnn}
\end{equation}
\end{widetext}

To simplify the integration, we introduce a coherence length $\lambda_c$, where the mass fluctuates coherently only inside the volume $(\lambda_c/2)^3$, and incoherently otherwise. The $\lambda_c$ is given by the acoustic wavelength, $\lambda_c= v_s/f$.

The calculation of the integral is quite lengthy, we will consider two useful cases. The first one is called the short wavelength limit, $L \gg \lambda_c$. The result of the NN is simple,
\begin{equation}
\tilde{h}^2_{NNs}(\omega)= \frac{896\pi G^2\Delta\tilde{M}^2(\omega)}{3 \omega^4 L^2 \lambda_c^4}\, .
\label{eqnn1}
\end{equation}

The second case is the long wavelength limit, $L \ll \lambda_c$. One can obtain
\begin{equation}
\tilde{h}^2_{NNl}(\omega)= \frac{32768\pi G^2\Delta\tilde{M}^2(\omega)}{15\omega^4 \lambda_c^6}\, ,
\label{eqnn2}
\end{equation}
where the NN is independent of the arm length $\emph{L}$.

The next step is to find out the mass fluctuation spectrum, $\Delta\tilde{M}^2(\omega)$. According to the model \cite{saulson1984}, the mass fluctuation due to the atmospheric pressure fluctuation is
\begin{equation}
\Delta\tilde{M}^2_{air}(\omega) = \frac{1}{2}(\pi v_{w})^6 \frac{\rho_{air}^2}{p_{air}^2}\frac{\Delta\tilde{p}^2_{air}(\omega)}{\omega^6}\, ,
\end{equation}
where, $v_{w}\simeq 20 \, {\rm m/s}$ is the wind velocity, $\rho_{air} \simeq 1.3 \, {\rm kg/m^3}$ is the air density, $p_{air} \simeq 10^5 \, {\rm Pa}$ is the mean air pressure, and $\Delta\tilde{p}^2_{air}(\omega)$ is the atmospheric pressure fluctuation spectrum.

The mass fluctuations due to the seismic displacement is
\begin{equation}
\Delta\tilde{M}^2_{seism}(\omega) = \pi^6 \rho_e^2 v_{L}^4 \frac{\Delta\tilde{X}^2_{seism}(\omega)}{\omega^4}\, ,
\end{equation}
where $v_{L}\simeq 5 \, {\rm km/s}$ is the typical velocity of longitudinal seismic waves, $\rho_{e} \simeq 2.7\times 10^3 \, {\rm kg/m^3}$ is the mean density of the medium around the lab. $\Delta\tilde{X}^2_{seism}(\omega)$ is the seismic displacement noise spectrum.

To estimate the INN and SNN for ZAIGA-GW, we need to measure $\Delta\tilde{p}^2_{air}(\omega)$  and $\Delta\tilde{X}^2_{seism}(\omega)$. Currently, we have not done such measurements. For the atmospheric pressure fluctuation spectrum, we use the same estimation as \cite{saulson1984},
\begin{equation*}
\Delta\tilde{p}^2_{air}(\omega)\simeq \frac{3\times 10^{-6}}{(f/{\rm Hz})^2}\, {\rm Pa}^2\, {\rm Hz}^{-1}.
\end{equation*}

For the seismic displacement noise spectrum, we use the data from KAGRA \cite{kagra2018},
\begin{equation*}
|\Delta\tilde{X}_{seism}(\omega)| \simeq \frac{10^{-9}{\rm m}}{(f/{\rm Hz})^2} \, {\rm  Hz^{-1/2}} \, .
\end{equation*}

\begin{figure}[htpb]
\includegraphics[width=8.0cm]{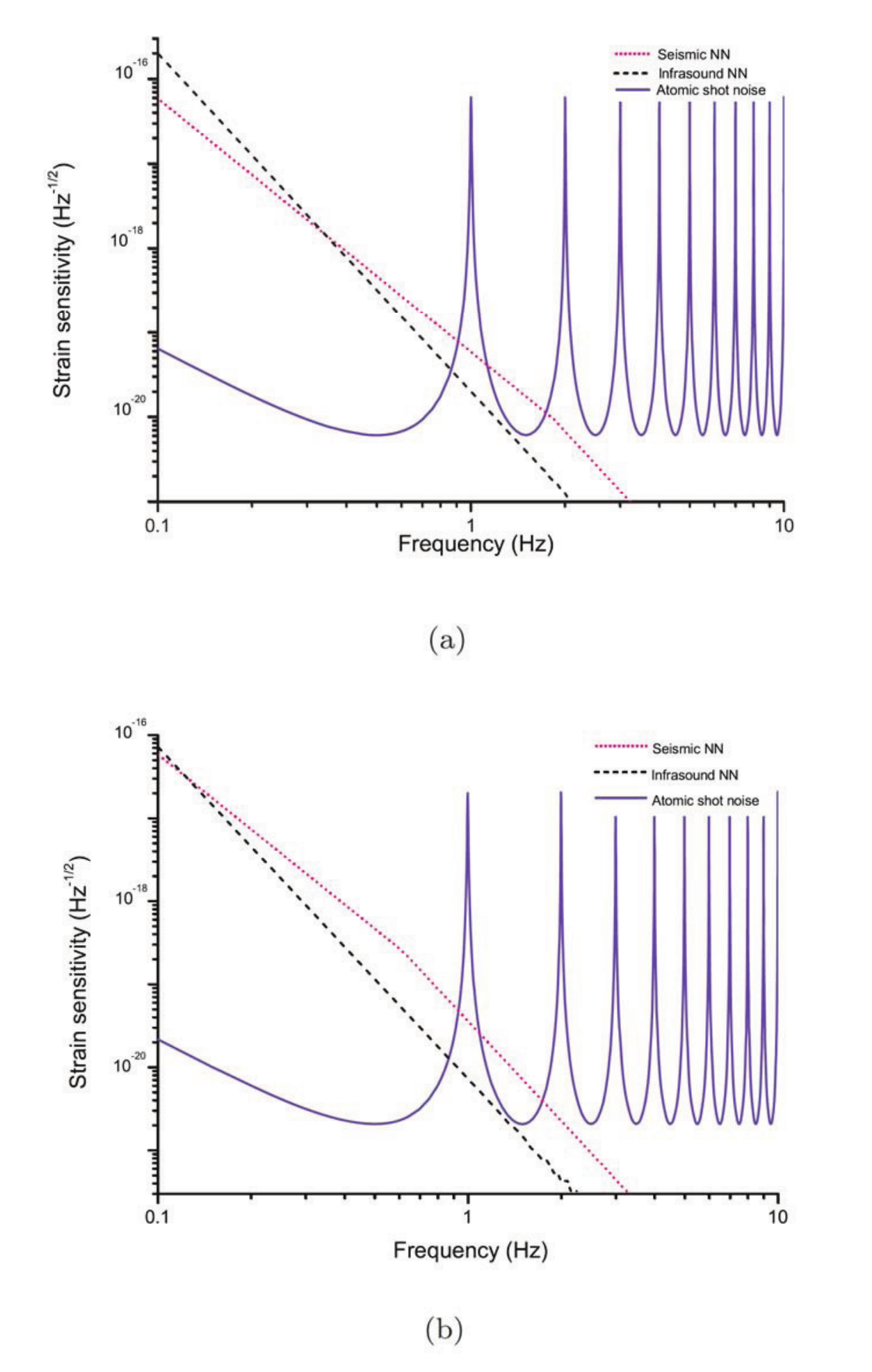}
\caption{Strain sensitivity of ZAIGA-GW. (a) For $L$ = 1 km. (b) For $L$ = 3 km. \label{f2.4}}
\end{figure}

Using Eqs. (\ref{eqnn1}) and (\ref{eqnn2}), we draw the PSD curves for INN and SNN in Fig.~\ref{f2.4}. It is clear that the Newtonian noise is very significant, and overtakes the shot noise below 1 Hz. Although, with the more refined models \cite{harms2015}, our numerical results may be revised by small factors, the main features of our PSD curves would not change.

\subsubsection{2.2.3 Technical requirement of ZAIGA-GW}

After the evaluation of main limits on the detection sensitivity, we now want to discuss the technical requirements of ZAIGA-GW to achieve the desired sensitivity. The following points need to be taken care of.

$\bullet$ The size of AIs. The height of AIs is assumed to be 5 meters, i.e. $T$=1 s, in the previous discussion. Since the 1/$T$ determines the frequency range of the GW detector, we want $T$ to be as large as possible to go into the lower frequencies. On the other hand, the larger the height is, the worse the interference fringe contrast is. We have to make a suitable balance between the height of AIs and the frequency range.

$\bullet$ The choice of atom species. We assume the flux intensity of $10^{14}$ atoms/s for $^{87}$Rb atoms to get a strain sensitivity $< 10^{-20}/\sqrt{{\rm Hz}}$ in the middle frequency band (0.1$\sim$10 Hz). The state-of-art atom flux intensity is about $10^{8}$ atoms/s \cite{treutlein2001}. It is a big challenge to increase the atom flux intensity. Of course, the requirement can be relaxed, if entangled atomic states could be used. For example, the atom flux intensity could be lowered to $10^{12}$ atoms/s with a 20 dB reduction in the detection phase noise \cite{chaibi2016}. Further discussion on preparing high-intensity atomic beams is given in Sec.5.

$\bullet$ The large momentum transfer. ZAIGA-GW requires the LMT to be of order $N$=1000. The 102 $\hbar k$ LMT has been demonstrated for $^{87}$Rb atoms \cite{chiow2011}. The authors claimed that, with technical improvements, there seemed no impediments to higher LMT, perhaps even exceeding 1000 $\hbar k$. The technical difficulties are how to make better wave front quality, brighter atom sources, and higher pulse efficiency.

$\bullet$ The laser noise. GWs induce the strain effect along the path of light beams. This effect is then transferred to atoms during their interactions with lights from the passive laser. Thus, noise from the passive laser goes into the phase shift of AIs. According to the estimation \cite{kasevich2008}, the laser frequency noise is roughly $\delta k L$. If we require this noise to be smaller than the shot noise, the required laser frequency stability is $10^{-16}$, which can be achieved by locking the laser to high finesse cavities \cite{ye2012}. Actually, this requirement can be relaxed for ZAIGA-GW since it consists of three detection baselines in a triangle configuration. The noise from the passive laser is common between any two baselines, thus, it is greatly reduced by taking differential measurements.

$\bullet$ The vibration isolation for the optics. Since the atoms are in free fall, the vibration isolation requirement for the GW detector based on laser-linked AIs is less stringent, which is one of the advantages compared to laser interferometer GW detectors. We estimate the required level of vibration isolation as Chaibi \emph{et al}. \cite{chaibi2016}. The vibrational noise goes into the GW detector through its effect on the positions of the lasers and beam splitting mirrors. The position noise results in a frequency noise for the light beams, $\tilde{\delta x}\, k_{l} f$. In the end, this frequency noise yields a phase noise in AIs, $k_{{\rm eff}} L 2\pi \tilde{\delta x} f/c$. If we require this phase noise to be smaller than shot noise, then we have $\tilde{\delta x} < \frac{10^{-14}{\rm m}}{(f/{\rm Hz})} \, {\rm Hz^{-1/2}}$ in the 0.1$\sim$10 Hz range. Thus, the vibration isolation requirement for ZAIGA-GW is about five orders of magnitude. Below  a couple of Hertz, vibration isolation is very difficult, which needs further study.

$\bullet$ The NN subtraction. From our rough estimation, we have found that the NN remains very significant below 1 Hz. To mitigate the NN, the first and important step is choosing very quiet, underground sites. But that is not enough. As suggested and outlined \cite{harms2015, et2011a}, additional suppression of the NN can be obtained by using dedicated arrays of auxiliary sensors (such as seismometers and microphones) to monitor density fluctuations near the AIs. One can then compute the real-time NN, and subtract it from the detection data. However, this is a very complicated job, and we need further detailed investigation.

\subsection{2.3 GW sources for ZAIGA-GW}

The ZAIGA-GW is targeting to detect the GW in frequency band 0.1$\sim$10 Hz . In this band, the most promising source is the black hole (BH) binaries. The GW radiation from compact binaries, with component mass $m_1$ and $m_2$, can be well calculated in the quadrupole approximation \cite{peters1964, allen2012}. Our first step is to classify the population of BH binaries which emit GWs in the ZAIGA-GW band. We calculate the final frequency of the BH binary coalescence and the duration of GW radiation from 0.1 Hz to 10 Hz or merger (if the merger happens frequency band 0.1$\sim$10 Hz ). The masses of BH binaries are chosen in the region of $[10^2, 10^5] M_{\odot}\times[10^2, 10^5] M_{\odot}$ and the mass ratio $m_1/m_2$ is limited in [1, 100], where $M_{\odot}$ denotes the solar mass, and $m_1 \geq m_2$ is assumed. The EOBNR (effective-one-body with numerical-relativity) approximant in LSC Algorithm Library Suite (LALSuite) is used to do this investigation \cite{bohe2017}. In our calculation, we neglect the BH spin effect.

\begin{figure}[htpb]
\includegraphics[width=8.0cm]{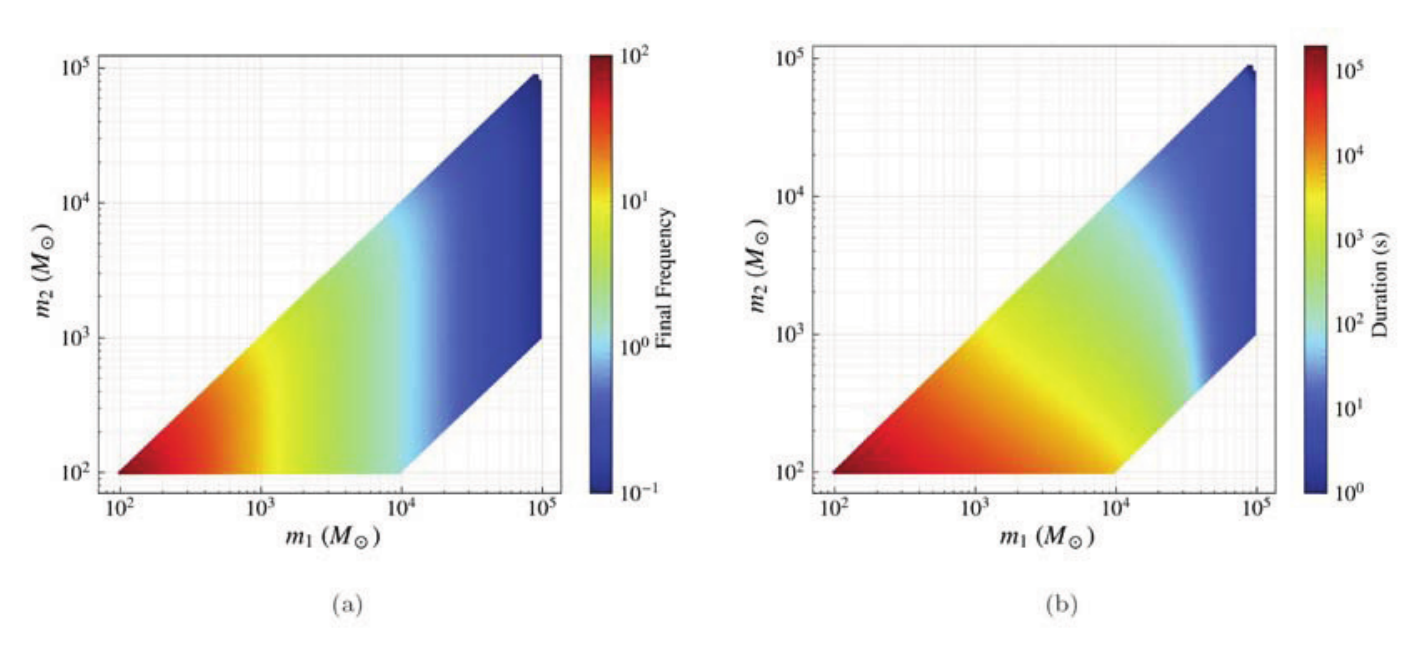}
\caption{The frequency and duration of the BH binaries. (a) The final frequency of the IMBH binaries. (b) The duration of the BH binaries from 0.1 Hz to the final frequency. \label{f2.5}}
\end{figure}

The final frequency of such BH binaries is shown in Fig.~\ref{f2.5}(a). As we can see from the plot, the BH binaries in the $10^2 M_{\odot}-10^5 M_{\odot}$ mass range (also called the intermediate-mass black hole (IMBH) binaries) can emit GWs in the ZAIGA-GW band.

The duration $T_{{\rm chirp}}$ (also called chirp time) of a BH binary at frequency $f$ is defined to be the remaining lifetime from the moment of emitting GWs of frequency $f$ to the final merger,
\begin{widetext}
\begin{equation}
T_{{\rm chirp}}= \frac{5}{256\eta}\frac{G M_{tot}}{c^3} \left\{ \nu^{-8}+\left( \frac{743}{252}+\frac{11}{3}\eta\right)\nu^{-6}-\frac{32}{5}\pi\nu^{-5}+\left(\frac{3058673}{508032}+\frac{5429}{504}\eta +\frac{617}{72}\eta^2 \right)\nu^{-4}\right\} \,
\label{tchirp}
\end{equation}
\end{widetext}
where
\begin{equation}
\nu=\left(\frac{\pi G M_{tot}}{c^3}f \right)^{1/3}\, .
\end{equation}

We introduce the total mass $M_{tot} = m_1 + m_2$, the chirp mass $M_{ch} = (m_1 m_2)^{3/5}/M_{tot}^{1/5}$, and the symmetric mass ratio $\eta = m_1 m_2/M_{tot}^2$. The result is depicted in Fig.~\ref{f2.5}(b). For the BH binaries with mass as low as a few hundreds $M_{\odot}$, the duration in our targeting band is more than 1 day. While, for the heave BH binaries, the duration is short as several seconds.

To illuminate the response of a detector to a GW signal, we choose a binary system in spiral time domain waveform as follows \cite{maggiore2007}
\begin{eqnarray}
h_+(t) &=& \frac{4}{D}\left(\frac{G M_{ch}}{c^2}\right)^{5/3}\left(\frac{\pi f}{c}\right)^{2/3}\frac{1+ {\rm cos}^2 \iota}{2}\, {\rm cos}\,\Phi (t)\nonumber \\
h_{\times}(t) &=& \frac{4}{D}\left(\frac{G M_{ch}}{c^2}\right)^{5/3}\left(\frac{\pi f}{c}\right)^{2/3}{\rm cos}^2 \iota\, {\rm sin}\,\Phi (t)\, ,
\label{eqh12}
\end{eqnarray}
where $D$ is the luminosity distance, $\iota$ is the inclination of binary system, and $\Psi$ is the phase of the waveform. As we can see from Eq. (\ref{eqh12}), the amplitude of the GW is inversely proportional to $D$, and mainly depends on the chirp mass. The amplitude difference of the two polarization depends on the inclination angle. The GW strain measured by a detector is
\begin{equation}
h=F_+(\theta,\phi,\psi)h_+ + F_{\times}(\theta,\phi,\psi)h_{\times},
\end{equation}
where $F_+$ and $F_{\times}$ are the antenna pattern functions of one detector, which depend on the relative angles between the detector and source location ($\theta,\phi$), and GW polarization angle $\psi$. A frequency domain waveform is beneficial for the matched filter calculation, which is used in the subsequent of this section. The stationary phase approximation inspiral waveform is \cite{allen2012}
\begin{widetext}
\begin{equation}
\tilde{h}(f) = -\left(\frac{5}{24\pi}\right)^{1/2}\left(\frac{G M_{\odot}}{c^2 D_{eff}}\right)\left(\frac{\pi G M_{\odot}}{c^3}\right)^{-1/6}\left(\frac{M_{ch}}{M_{\odot}}\right)^{5/6}f^{-7/6}{\rm exp}[-i \Psi(f)]\, ,
\label{eqh}
\end{equation}
\end{widetext}
where $\Psi(f)$ is the phase of the waveform depends on the frequency, total mass, and mass ratio etc. $D_{eff}$ is the effective distance, defined as
\begin{equation}
D_{eff}=D\left[\left(\frac{1+ {\rm cos}^2 \iota}{2}\right)^2 F_+^2 + {\rm cos}^2 \iota \, F_{\times}^2 \right] .
\end{equation}

Then we can use the matched filter algorithm to calculate signal-to-noise ratio (SNR) for a compact binary coalescence
\begin{equation}
\rho^2=4\int_{f_{{\rm low}}}^{f_{{\rm high}}}\frac{|\tilde{h}(f)|^2}{S_n(f)}df\, ,
\end{equation}
where $f_{low}$ and $f_{high}$ are the low and high frequency cutoff, respectively. $S_n(f)$ is the detector power spectral density, and $\tilde{h}(f)$ is the GW waveform in frequency domain. However, as we investigated above, for the lighter binary BH systems merger at frequency higher than 10 Hz, their GW waveforms could be approximately described
by the Eq. (\ref{eqh}), while for the heaver BH binary merger in the ZAIGA frequency band, the inspiral-merger-ringdown (IMR) waveform is required. The IMR Phenom approximant is employed to simulate GW waveform which includes binary IMR \cite{khan2016}.

To estimate the ZAIGA-GW detectability to the IMBH binaries, we calculate the SNR for the GW signals from different mass binary system at different distance. And we only consider the optimal case which the sources are optimally located and oriented. In this case, the effective distance $D_{eff}$ is equal to the true luminosity distance $D$. The corresponding SNR is named as optimal SNR which will be the best situation in reality.

\begin{figure}[htpb]
\includegraphics[width=8.0cm]{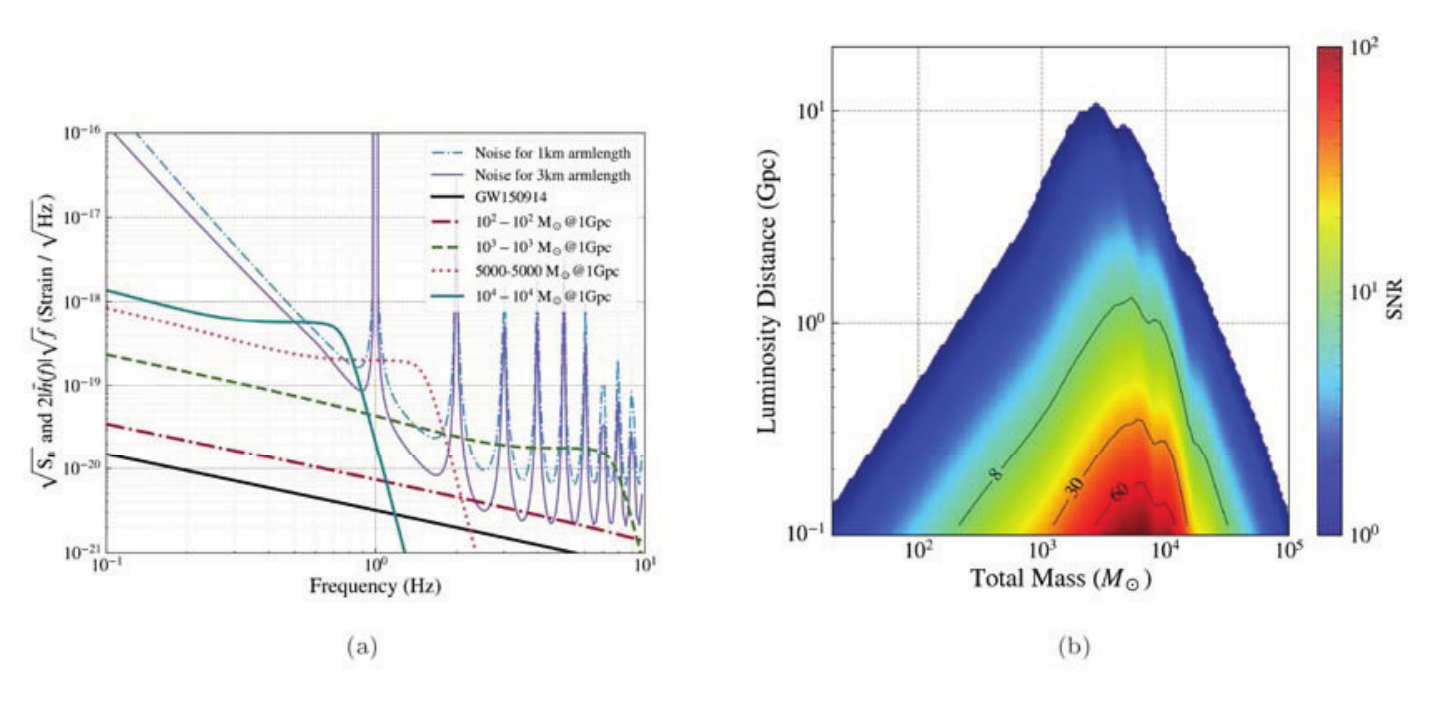}
\caption{The raw noise budget and optimal SNR distribution for ZAIGA-GW. (a) The raw noise budget for ZAIGA-GW with 3 km arm length configuration and the potential detectable BH binary merger in the sensitive band. (b) The optimal SNR distribution for equal mass BH binaries at different luminosity distance for single ZAIGA-GW detector. \label{f2.6}}
\end{figure}

With the raw noise budget for ZAIGA-GW, we depict several strain curves for some typical BH binary systems in Fig.~\ref{f2.6}(a). For comparison, the curve for GW150914 in the ZAIGA-GW frequency band is also shown, and it is just out of reach of raw ZAIGA-GW. In Fig.~\ref{f2.6}(b), it's shown that the corresponding SNR distribution for different total mass (with mass ratio equal to 1) BH binaries at different distance. The most promising source would be $\sim 5000-5000 M_{\odot}$ which we can reach the 1 Gpc assuming the SNR threshold equal to 8.

\begin{figure}[htpb]
\includegraphics[width=8.0cm]{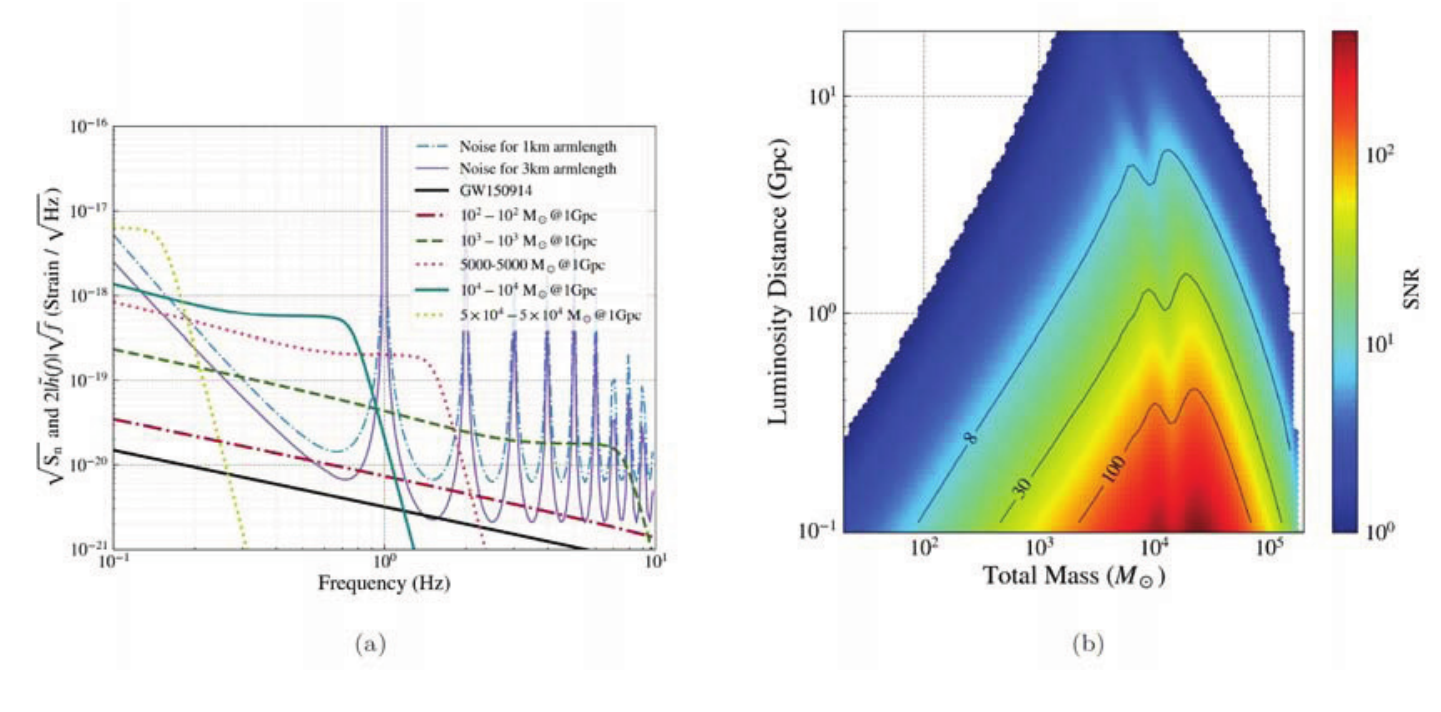}
\caption{The optimal noise budget and optimal SNR distribution for ZAIGA-GW. (a) The optimal noise budget (50 times NN subtraction assumed) for ZAIGA with 3 km arm length configuration and the potential detectable BH binary merger in the sensitive band. (b) The optimal SNR distribution for equal mass BBH at different luminosity distance for single ZAIGA-GW detector. \label{f2.7}}
\end{figure}

Suppose that there could be a factor of 50 Newtonian noise subtraction for ZAIGA-GW, which is also suggested for ET \cite{ET2010}. We call it optimal ZAIGA-GW configuration, and the corresponding curves are shown in Fig.~\ref{f2.7}. In this case, ZAIGA-GW is most sensitive to the binary source $\sim 10^4-10^4 M_{\odot}$ which we can reach the 5 Gpc with the SNR threshold equal to 8.

Due to the IMBH mass distribution and formation process are unknown, the merger rate of the IMBH binary is quite uncertain. Based on the merger rates of IMBH calculated in \cite{matsubayashi2004}, we estimated the potential detection rates by ZAIGA-GW. By one year observation with the optimistic configuration as shown in Fig.~\ref{f2.7}, ZAIGA-GW may have several IMBH detections per year.

\begin{figure}[htpb]
\includegraphics[width=8.0cm]{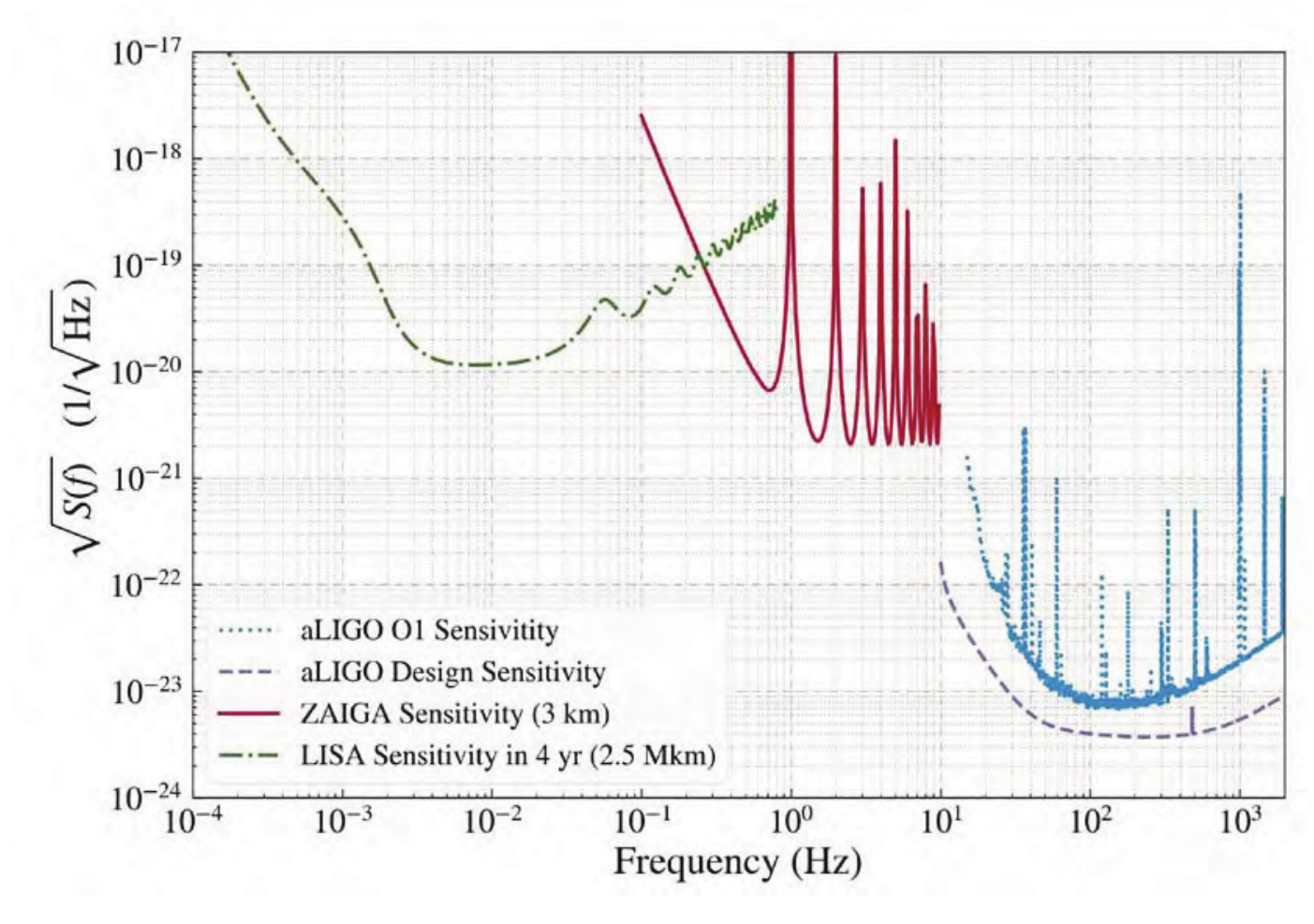}
\caption{Comparison of our proposed ZAIGA with LIGO and LISA. \label{f2.8}}
\end{figure}

The ZAIGA-GW is an underground GW detection plan based on laser-linked AIs, in an equilateral triangle configuration. It has several advantages: the generation of null data stream, a full detection of GW polarizations, and $100\%$ on duty. The ZAIGA-GW can fill in the detection gap between the ground-based laser interferometer GW detectors (such as LIGO) and the future space-based GW detectors (such as LISA), as shown in Fig.~\ref{f2.8}. Although some plausible candidates for intermediate-mass black holes (IMBH) have been reported \cite{fabbiano2005, mezcua2018}, no direct evidence has been found. As shown in Fig.~\ref{f2.5}(a), the final frequencies of IMBH binaries are out of LIGO's sensitive band, LIGO could not detect them. On the other hand, with the optimal sensitivity $< 10^{-20}/\sqrt{{\rm Hz}}$ in the middle frequency band, ZAIGA-GW has a chance to detect the GWs radiated from IMBH binaries up to the distance of 10 Gpc.

The other notable thing is that stellar-mass BH binaries (${\rm mass} \sim 100 M_{\odot}$) in the middle frequency band, up to the distance of 1 Gpc, are also detectable for optimal ZAIGA-GW. After emitting in the middle frequency band, these sources step into the LIGO band. Since they spend much longer time in the middle frequency band, ZAIGA-GW can send early notices to the ground-based laser interferometer GW detectors, and finally join together to provide a multi-band detection of these sources.

The ZAIGA-GW is extendable. The site is so large that it is possible to install laser interferometer GW detectors, together with the detectors based on laser-linked AIs. With the laser interferometer GW detectors targeting at the high frequency band, and the AI-based detectors targeting at the middle frequency band, we could finally have a multi-band GW detection plan.

\section{3. ZAIGA-EP: Equivalence Principle Test}

The ZAIGA-EP apparatus will locate at the 300-meter-vertical tunnel, where two 10-meter atom fountains mounted on both top and bottom of the tunnel. The primary goal of ZAIGA-EP is to test the weak equivalence principle using AIs. The secondary goals are searching for spin-gravity coupling\cite{Rosi2017}, investigating quantum mechanical superposition on macroscopic scales\cite{Kovachy2015} and 300-meter-vertical prototype antenna for mid-frequency gravitational waves.

\subsection{3.1 Equivalence principle}

Gravity is one of the four fundamental interaction forces of nature, all attempts to quantize gravity and to unify it with the other three forces (strong interaction, weak interaction, and electromagnetic interaction) suggest that the GR may not be the last word\cite{Will2014}. GR test has important scientific significance. One of the theoretical foundations of GR is Einstein's Equivalent Principle (EEP). The EEP consists of three parts: Weak Equivalence Principle (WEP, also known as Universality of Free Fall(UFF)), Local Lorentz Invariance(LLI) and Local Position Invariance(LPI). The WEP is the core content of the EEP. The WEP has many equivalent expressions, and the most common expression is that the trajectory of a free fall object does not depend on its internal structure and composition. Many experiments to date have proved that the WEP is correct within a certain precision. However, almost all new theories that try to unify the gravity and the standard model (such as string theory, loop quantum gravity theory, extra dimensional theory, non-commutative geometry, and fifth force) require WEP to be broken\cite{Will2014}. To validate these new theories and explore the applicable scope of WEP, more accurate WEP tests are needed. The accuracy of WEP test using macroscopic objects \cite{Dittus1996,Wagner2012,Williams2012,Touboul2017} has reached a level of $10^{-13} \sim 10^{-15}$. The early work of WEP test using micro-particles was done by a neutron interferometer\cite{Koester1976,Bertolami2003}, the accuracy is only $10^{-4}$ due to experimental technical limitations. In recent years, the rapid development of atom interferometry \cite{Kasevich1991,cronin2009,WangJ2009,WangJ2015} provides a new way to test WEP by using micro-particles. The AI use quantum systems (atoms) to measure gravity, this directly links quantum mechanics with GR, which is helpful to provide clues for combining the two theories. The comparation measurement between an atom gravimeter and a laser interference absolute gravimeter (FG-5) \cite{Peters1999,Merlet2010} is a WEP test using macro-objects and micro-particles. By measuring and comparing the gravitational accelerations of different atoms with atom interferometry, the WEP of the microscopic particles can be tested.

\subsection{3.2 Equivalence principle test using microscopic particles}

In recent years, physicists carried out the micro-particle-based WEP tests using different AIs\cite{Fray2004,Bonnin2013,Schlippert2014,Tarallo2014,zhan2015,Barrett2016,Duan2016,Rosi2017}, the precision of E\"{o}tv\"{o}s coefficient, $\eta$, is between $10^{-4}\sim 10^{-9}$. Fray $\emph{et al}$.\cite{Fray2004} from Max Planck Institute for Quantum Optics (MPQ) conducted the first AI-based WEP test, and the obtained E\"{o}tv\"{o}s coefficient is $\eta = (1.2\pm 1.7) \times 10^{-7}$. Bonnin $\emph{et al}$. \cite{Bonnin2013} from National Office for Studies and Aerospace Research (ONERA) tested WEP using $^{85}$Rb-$^{87}$Rb dual atom interferometers, the obtained E\"{o}tv\"{o}s coefficient is $\eta = (1.2 \pm 3.2) \times 10^{-7}$. Schlippert $\emph{et al}$.\cite{Schlippert2014} from Leibnitz University Hannover (LUH), tested WEP using non-isotopic $^{87}$Rb-$^{39}$K atom interferometers, and $\eta$ = $(0.3 \pm 5.4) \times 10^{-7}$. Tarallo $\emph{et al}$.\cite{Tarallo2014} from University of Florence (UF), conducted a WEP test using Fermion isotope $^{87}$Sr and Boson isotope $^{88}$Sr atoms in a optical lattice, the result is $\eta = (0.2 \pm 1.6) \times 10^{-7}$. Zhou $\emph{et al}$.\cite{zhan2015} from Wuhan Institute of Physics and Mathematics(WIPM) realized WEP test using the four-wave double-diffraction Raman-transition (FWDR) $^{85}$Rb-$^{87}$Rb atom interferometers, the result is $\eta = (2.8 \pm 3.0) \times 10^{-8}$. Barrett $\emph{et al}$.\cite{Barrett2016} from University of Bordeaux (UB) carried out WEP test using $^{87}$Rb-$^{39}$K atoms in microgravity conditions of an aircraft during parabolic flight, the measured E\"{o}tv\"{o}s coefficient in 0-$\emph{g}$ environment is $\eta = (0.9 \pm 3.0) \times 10^{-4}$. This experiment is a verification for the key technologies of the satellite-borne AI-based WEP test project "STE-QUEST"\cite{Altschul2015,Aguilera2014}. Duan $\emph{et al}$. \cite{Duan2016} from Huazhong University of Science and Technology (HUST) completed the UFF test using different spin-orienting atoms ($^{87}$Rb, $m_{F}= \pm1$), the result is $\eta = (0.2 \pm 1.2) \times 10^{-7}$. Rosi $\emph{et al}$.\cite{Rosi2017} form University of Florence (UF) conducted a EP test using superposition state atoms, the result is $\eta = (3.3 \pm 2.9) \times 10^{-9}$.

The sensitivity of an AI depends on the free evolution time of the atoms involved in the interference process. Long-baseline AI is an effective way to increase the free evolution time of atoms. For this purpose, several groups around the world are building or going to build long-baseline AIs. The maximum launch height of atoms in the 10-meter atom fountain developed by Kasevich's group at Stanford University is 9 m \cite{Dickerson2013,Kovachy2015},  where the effective interference area inside the magnetic shield is 8.2 m, and the longest free evolution time can reach 1.34 s \cite{Dimopoulos2007,Dimopoulos2008}. The very long baseline dual Rb-Yb AI, designed by Rasel's group \cite{Hartwig2015} at LUH, has a height of 10 m and an effective interference area of 9 m. The AI array for gravitational wave detection (MIGA) \cite{chaibi2016} is being built by Bouyer's group at UB. Zhan's group at WIPM designed and developed a 10-meter AI \cite{ZhouL2011} for high-precision WEP test, they obtained the time-of-flight signal of 12-m-height atom fountain recently. To further improve the accuracy of the AI-based WEP test, Ksevich's group is planning to construct a 100-m large AI \cite{Dimopoulos2007}. The high-precision WEP tests using long-baseline AIs may lead to the revision of GR or the support to theory of quantum gravity.

\subsection{3.3 Design of ZAIGA-EP}

\subsubsection{3.3.1 General principle}
The AI is a precision instrument that is sensitive to many physical parameters, and the gravity field is one of them. The phase shift of an AI is related to the gravitational acceleration of atoms, $\emph{g}$, the effective wave vector of Raman laser, $\emph{k}_{eff}$, and the square of the time interval of the Raman laser, $T^{2}$. The gravitational acceleration sensed by the atom can be extracted by the phase shift of the interference fringe. Since $\emph{k}_{eff}$, \emph{T} can be precisely controlled, if the phase shift of the atom interference fringes can be accurately measured, the precision measurement of gravity can be achieved. The WEP test experiment is usually to measure the tiny acceleration difference between two objects with different materials in the gravitational field. If the gravitational acceleration of different isotope atoms (e.g., $^{85}$Rb and $^{87}$Rb) are measured, then WEP can be tested by microscopic particles. The E\"{o}tv\"{o}s coefficient is expressed as,
\begin{equation}
\eta = \frac{g_{87}-g_{85}}{(g_{87}-g_{85})/2},
\label{eqn3.1}
\end{equation}
where, $g_{85}$ and $g_{87}$ are gravitational acceleration of $^{85}$Rb and $^{87}$Rb atoms, respectively. The main parameters affecting the measurement accuracy of the AI are the velocity of atoms, $\upsilon$, (corresponding to the temperature of atoms), the free fall time of atoms, \emph{T}, the laser frequency, $\omega$, and the height of atom fountain, \emph{h}. These parameters are controllable. Consider a 300-meter-high atom fountain, the cold atom can fall in the gravitational field for 7.7 s. By controlling the experimental parameters, the paths of the two atoms can be overlapped as much as possible, then some noise (such as gravity gradient, vibration noise, phase noise between Raman laser, etc.) are common mode for the two species atoms to some extent, they can be suppressed commonly.

\subsubsection{3.3.2 The overall structure of ZAIGA-EP}

\begin{figure}[pb]
\includegraphics[width=8.0cm]{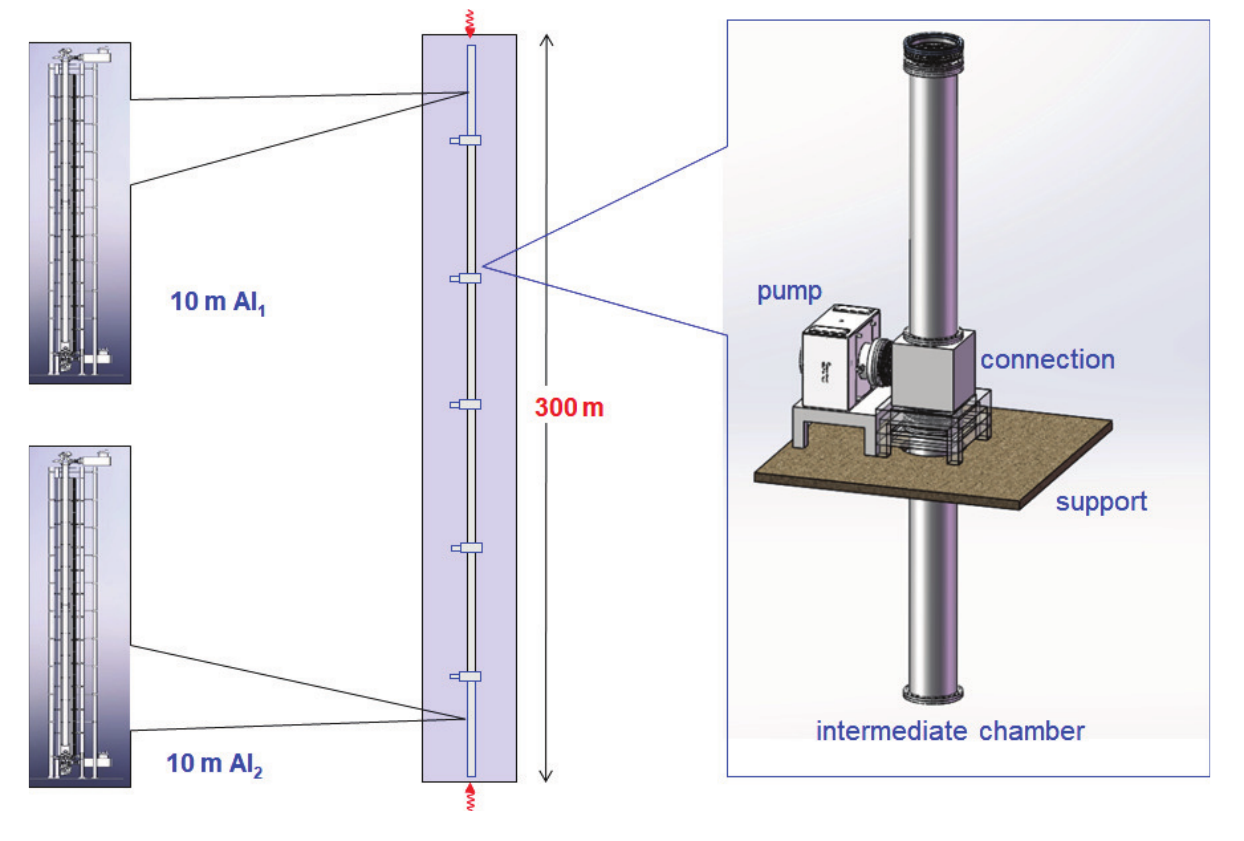}
\caption{Schematic diagram of vacuum system for 10-meter AI system. \label{f3.2}}
\end{figure}

The ZAIGA-EP apparatus includes two sets of 10-meter AIs and a 300-meter-vertical high-vacuum tube chamber. Where, two AIs are mounted on both top and bottom of the 300-meter-vertical chamber, the 300-meter chamber is for transmitting link lasers between two AIs. To maintain a high-vacuum environment within the 300-meter tube chamber, a set of ion-pump system need to be installed every 50 meters. The overall structure of the atomic interferometer is shown in Fig.~\ref{f3.2}. Each set of 10-meter AI consists of a two dimensional magneto-optical trap (2D-MOT), a three dimensional magneto-optical trap (3D-MOT), fountain chamber, several vacuum pumps, a vibration isolation system, a magnetic shielding system, a signal detection system, a timing control system, lasers and optical system. Among them, the 2D-MOT and the 3D-MOT are used to prepare cold atom clouds, the fountain chamber is for free fall and interference of atoms, the detection window is used to acquire the interference signal, the magnetic shielding system provides a low background uniform magnetic field for the interference zone. The vertical height of the 10-meter AI apparatus is 12 m, and the height of the magnetic shielding area is 10 m.

The 10-meter AI can prepare two different species cold atoms for the microparticle WEP test. The upper 10-AI works in drop mode, where cold atom clouds have a free fall distance of 10 m and a maximum free fall time of 1.4 s inside the magnetic shielding, thus the free evolution time in the Mach-Zehnder configuration AI is $\emph{T}$ =0.5$\sim$0.7 s. The lower 10-AI works in launching mode, where cold atom clouds have a free fall distance of 20 m and a maximum free fall time of 2.8 s inside the magnetic shielding, thus the free evolution time in the Mach-Zehnder configuration AI is $\emph{T}$ =1.0$\sim$1.4 s. In the far future, it is also possible to use all the 300-meter vertical tube chamber, so that the cold atom cloud prepared by the upper MOT can freely fall for 7.7 s.


\subsubsection{3.3.3 Main technical parameters and expected uncertainty}
Currently, the uncertainty of $\eta$ measure by terrestrial AI is 10$^{-8}$ ($^{85}$Rb-$^{87}$Rb) \cite{zhan2015} and 10$^{-9}$ ($^{87}$Rb internal superposition) \cite{Rosi2017}. To further improve the WEP test accuracy, besides to prolonging the free evolution time \emph{T}, other measures are needed, including further increasing the number of atoms involved in the interference process to reduce the quantum projection noise, using ultra-cold atoms to reduce the divergence of atom clouds during their free fall process, using large momentum transfer to increase the interference loop area, isolating background vibration to reduce vibrational noise; shielding residual magnetic field to provide a low background uniform magnetic field for interference region, compensation of Coriolis effect; using Bragg transitions appropriately to reduce the magnetic sensitivity of AIs, the use of single photon transition mechanism of alkaline earth metal elements (such as strontium) or lanthanide (such as ytterbium) to reduce the magnetic shielding requirements of long-baseline AI, finally, to use entangled atom source (such as superposition, inter species entanglement) to achieve the Heisenberg limit. The main technical parameters of ZAIGA-EP are listed in Table~\ref{ta1}, where the frequency band of vibration background is  0.01$\sim$10 Hz.

\begin{table*}[htbp]
\caption{The main technical parameters of ZAIGA-EP.}
{\begin{tabular}{@{}lccc@{}}
\toprule
Parameters & Near future drop mode & Near future launch mode & Far future drop mode \\ \colrule
Atomic species & $^{85}$Rb-$^{87}$Rb & $^{85}$Rb-$^{87}$Rb & $^{87}$Sr-$^{88}$Sr \\
Atom number & 10$^{6}$ atoms/s & 10$^{6}$ atoms/s & 10$^{6}$ atoms/s \\
Free fall time & 1.4 s & 2.8 s & 7.7 s \\
Momentum transfer & 2 $\hbar$$\emph{k}$ & 2 $\hbar$$\emph{k}$ & 200 $\hbar$$\emph{k}$ \\
Atom temperature & 1 $\mu$K & 1 $\mu$K & 50 nK \\
Vibration background & 10$^{-9}$g/$\sqrt{Hz}$ & 10$^{-9}$g/$\sqrt{Hz}$ & 10$^{-9}$g/$\sqrt{Hz}$ \\
Residual magnetic field & 10 nT & 10 nT & 500 nT \\
Transition type & Raman & Raman & Bragg \\
Noise limit & 1/$\sqrt{N}$ & 1/$\sqrt{N}$ & 1/\emph{N} \\
Expected uncertainty & $\eta$$\sim$10$^{-13}$ & $\eta$$\sim$10$^{-13}$ & $\eta$$\sim$10$^{-15}$ \\ \botrule
\end{tabular} \label{ta1}}
\end{table*}

\subsection{3.4 Key techniques for ZAIGA-EP}

There are many technical problems need to be overcome in the ZAIGA-EP equipment, the key techniques include long-baseline atom fountain, large-scale magnetic shielding, dual-species synchronous atom interference, active vibration isolation, wave-front phase noise suppression, Coriolis effect compensation, signal detection and data processing methods.

\subsubsection{3.4.1 Long-baseline atom interferometer}

\begin{figure}[htpb]
\includegraphics[width=8.0cm]{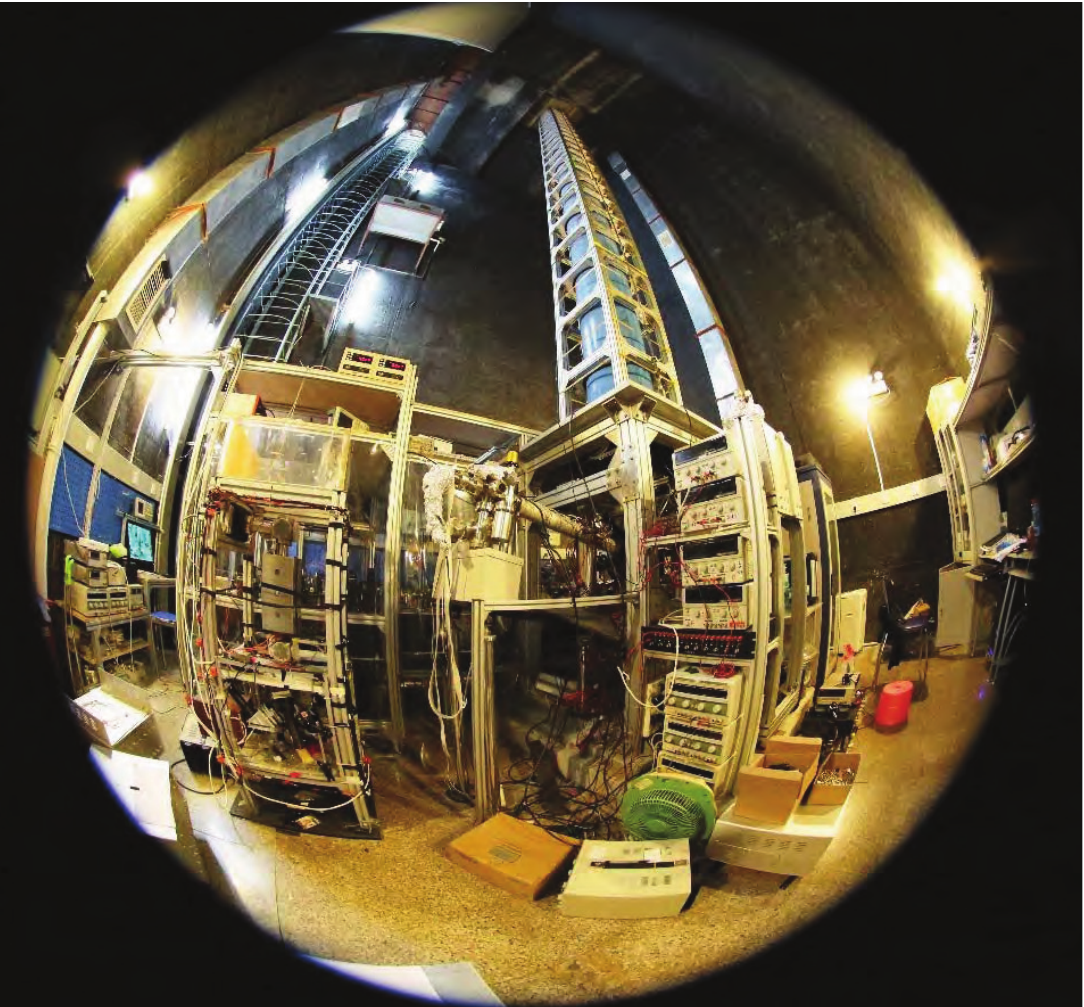}
\caption{Fisheye photo of the 10-m atom interferometer setup. \label{f3.5}}
\end{figure}

The core device of ZAIGA-EP is a long-baseline AI. To ensure that the atom interference zone reaches 300 m, we need to prepare enough cold atoms in the MOT and launch them over 300 m to form atom fountain. We developed a 10-meter AI in 2011, and obtained a 6-m-heigh fountain signal \cite{ZhouL2011}. The 10-meter AI consists of three parts: the bottom MOT, the fountain chamber, and the top MOT. The whole system is higher than 12 m, and the effective interference area inside the magnetic shield is 10 m. The fisheye photo of the 10-meter AI setup is shown as Fig. ~\ref{f3.5}. Recently, we improved the preparation efficiency of cold atoms by modifying the MOT chamber, the state preparation region and the detection region, and we observed the time of flight (TOF) signal of 12-meter-height fountain by lowering the temperature of atoms.


\subsubsection{3.4.2 Large-scale magnetic shielding}

When we developed the previous 10-meter AI, after several rounds of integral welding and annealing tests, we gradually overcome the bottleneck of the large-size annealing technique, developed the active magnetic compensation technology inside and outside shielding layer, and finally reduced the magnetic field fluctuations in the interference zone from 600 nT to 8 nT. The ZAIGA-EP needs a 300-meter large-scale magnetic shielding system, the magnetic field undulation in the interference region should be less than 10 nT, this a new technical challenge.

\subsubsection{3.4.3 Dual-species synchronous atom interference}

Common-mode measurement is an effective way to eliminate common-mode noise. In the traditional single-diffraction Raman transition atom interference process, the influence of gravity gradient cannot be completely suppressed in the dual-species AI due to the asymmetry interference path. The crosstalk between different frequency Raman lasers also introduces phase noise. L\'{e}v\`{e}que \emph{et al}. \cite{Leveque2009,Malossi2010}used a double-diffraction Raman transition technique to improve the phase noise of a single-species AI. Zhou \emph{et al}.\cite{zhan2015} applied the double-diffraction Raman transition technique to a dual-species AI, proposed and implemented the FWDR to solve the crosstalk of multi-frequency lasers in $^{85}$Rb-$^{87}$Rb dual-species AI, and to suppresses the common-mode noise, the uncertainty AI-based WEP test was improved from 10$^{-7}$ to 10$^{-8}$.

\subsubsection{3.4.5 Active vibration isolation}

Vibrational noise can be transmitted into the phase of the atom interference fringes through the Raman lasers' mirror. To reduce the vibrational phase noise, it is necessary to isolate the Raman lasers' mirror from the ambient vibration source. The AI is more sensitive to vibrations with a frequency lower than 1/\emph{T}. To effectively suppress the 0.01$\sim$10 Hz low-frequency vibration noise in the AI, Tang \emph{et al}. \cite{Tang2014} designed an active vibration isolation system for the Raman lasers' mirror. In this vibration system, an seismometer is used to measure vibration of the mirror mount, and a voice coil motor is used to suppress the vibration of the mount by FPGA digital feedback control. The system control algorithm and parameters are optimized to reduce the resonant frequency of the vibration isolation system to 0.015 Hz, and the feedback gain is improved to as high as 60 dB. With this active vibration isolation system, the background vibration noise is suppressed by two orders of magnitude, and the power spectral density of vibration noise in 0.01$\sim$10 Hz is decreased to 10$^{-9}g/\sqrt{{\rm Hz}}$ level. After the active vibration isolation system is applied to the AI, the short-term stability of the interference fringes is significantly improved.

\subsubsection{3.4.6 Wave-front phase noise suppression}

The wave-front aberration of the Raman beam causes phase noise in the atom interference signal. To suppress the phase noise of the wave-front distortion, Hu \emph{et al}. \cite{HuJG2017} proposed an extend-rate-selection scheme to suppress the wave-front aberration noise in WEP test using dual-species isotope and non-isotope AIs. The simulation based on reasonable experimental parameters shows that the wave-front aberration noise can be suppressed ten times with this scheme, and the standard deviation of the $\eta$ using $^{85}$Rb-$^{87}$Rb and $^{41}$K-$^{87}$Rb dual-species AIs are $1.3 \times 10^{-14}$ and $3.0 \times 10^{-13}$, respectively.

\subsubsection{3.4.7 Coriolis effect compensation}
We designed and implemented a rotation compensation system for Raman lasers' mirror, and compensated the Coriolis effect caused by the rotation of the Earth. We use a tip-tilt platform which could moves around two orthogonal axes and drives by four piezoelectric ceramic transducer (PZT) to support and precisely adjust the tilt angle of the mirror, so that the rotating of the mirror is in opposite direction and equal rate to that of the Earth, thereby compensating for the Coriolis effect on the free-falling atoms. After the rotation compensation, the amplitudes of the interference fringes of the $^{85}$Rb and $^{87}$Rb atoms are simultaneously modulated, and the phases of the two interference fringes are also modulated at the same time.

\subsubsection{3.4.8 Signal detection and data processing methods}

The usual normalized detection \cite{Rocco2014,Biedermann2009} requires more time, complicated detection procedures, and higher requirements for probe beam diameter. To simplify the detection process, reduce the amplitude noise, Song \emph{et al}.\cite{Song2016} proposed and implemented a simple normalized detection method, which uses the quenching fluorescence signal during the initial preparation to normalize the population. To reduce the error of differential phase fitting, Wang \emph{et al}.\cite{WangYP2016} proposed a combination scheme for extracting the differential phase of gravity gradient measurement. By modulating the magnetic field of one AI, then extracting the differential phase of dual AI using ellipse and linearity fitting, so that the small differential phase can be accurately extracted under a large noise environment. This method makes up for the deficiency of the ellipse fitting and the Bayesian statistical method, and it can be extended to the dual-species AI-based WEP test. The WEP violation factor between various elements is related to the number and type of neutrons that make up them. Compared to isotopic atoms, there is a larger WEP violation factor between non-isotopic atoms\cite{Damour1996}. Chen \emph{et al}.\cite{Chen2014} analyzed the vibrational noise in atom interference fringes, proposed a proportional scanning phase method based on the correlation of vibration noise of dual-species AI. This method can suppress the common-mode vibration noise, and the rejection ratio is up to 140 dB. In addition, the proportional scanning phase is beneficial to obtain the optimal unbiased estimation of the WEP violation factor, expand the candidate’s category of WEP test.

\section{4. ZAIGA-CE: Clock-based gravitational redshift measurement and GW detector prototype}
We will present two ways to use accurate optical clocks in the quiet underground ZAIGA facility. The first one is clock redshift experiment ZAIGA-CE-R. The second one is a prototype of clock-based GW detector ZAIGA-CE-GW. The precision and stability of optical clocks has taken great stride during last 20 years. The recent demonstrated systematic uncertainty of optical clock is below 10$^{-18}$ \cite{Brewer1902}, the measurement instability of optical clock is $3.2 \times 10^{-19}$ \cite{McGrew2018}, and the development to 10$^{-20}$ level is envisaged. A clock GW mission with laser link would then be a good possibility, after the third-generation ground-based GW detectors and first-generation space borne laser-interferometric GW detectors (using TDI) \cite{nibook,Ni2016}, especially with AU or larger arm length.

Gravitational redshift is predicted by Einstein's GR as a classical phenomenon. The gravitational redshift arises when light moves away from a static massive object, such as the Earth or the Sun. It describes the fact that clocks in a gravitational field tick slower in a point of view of a distant observer. That is, it refers to
the shift of wavelength of a photon to longer wavelength (the red side in an optical spectrum) when observed from a point in a lower gravitational field. The gravitational redshift is a simple consequence of EEP. Observing the gravitational redshift in the solar system is one of the classical tests of GR. Gravitational redshifts are an important effect in satellite-based navigation systems such as GPS. The ZAIGA-CE-R is designed for clock-based gravitational redshift measurement.

\subsection{4.1 Measure gravitational redshift by ZAIGA-CE-R}

The basic concept of redshift measurements is to synchronize a pair of clocks when they are located closely to one another, and move them to different elevations. The gravitational redshift will decrease the oscillation frequency of the lower clock relative to the higher one. When we bring the clocks together afterwards and compare the number of elapsed oscillations, there will be a measurable phase shift between them. With a simple formula, the frequency difference of the two clocks is related to the two different location potentials\cite{Kleppner1970},

\begin{equation}
\frac{\Delta f}{f}=\frac{\Delta U}{c^{2}}.
\label{eqn4.1}
\end{equation}

The UFF and LLI have been verified experimentally to accuracies of 10$^{-13}$ or better. The LPI requires the outcome of a non-gravitational experiment to be independent of where and when it is performed. In practice, the highest precision tests of local position invariance are measurements of the gravitational redshift: the frequency of a clock is measured as a function of location. The fractional frequency difference takes the form,
\begin{equation}
\frac{\Delta f}{f}=(1+\beta)\frac{\Delta U}{c^{2}},
\label{eqn4.2}
\end{equation}
where, $\beta$ denotes the degree of violation of the local position invariance, which is a clock-dependent parameter. From the formula, it's obvious that there will be no variations other than those caused by gravity, that is, the gravitational redshift, if EEP holds($\beta$ = 0)\cite{Muller2010}. A famous version, called Gravity Probe A experiment(GP-A), compared a hydrogen maser in rocket with a ground-based clock. It is confirmed Einstein's prediction with the accuracy of roughly 10\%. The accuracy of $\beta$ is measured smaller than 7 $\times$ 10$^{-5}$\cite{Vessot1980}. The GP-A result has been held for almost half century, until recently it's reported there is an improvement of factor 5.6 with passive H-maser in European global satellite navigation system. The European Space Agency(ESA) use data spanning 1008 days from two Galileo satellites, GSAT-0201 and GSAT-0202, which were accidentally delivered on elliptic orbits. Fractional deviation of the gravitational redshift from the prediction by general relativity is measured to be (0.19$\pm$2.48)$\times$10$^{-5}$ at 1 $\sigma$\cite{Delva2018}.

There is another way to test the general relativity based on the assumption that if two clocks of different internal structures move together through a gravitational potential, their frequency ratio must be constant. In 2018, a null test of general relativity based on a long term comparison of cesium fountain clock and hydrogen masers is pushed to a higher accuracy level of(2.2$\pm$2.5)$\times$10$^{-7}$.\cite{Ashby2018}
Owing to the long-term drifts of H masers, there isn't too much improvement for the uncertainty in the LPI parameter using H masers and Cs fountains. Due to the higher stability of the optical clocks at least two orders of magnitude better than microwave clocks, future improvements may benefit from comparisons among the optical clocks.

With the development of optical clock during the past decade, its accuracy has surpassed the cold atom fountain clock. Two pioneering breakthroughs have been exhibited to unprecedented stability of 1 part in 10$^{19}$ recently\cite{Marti2018,McGrew2018}. Owing to the much shorter averaging time, lattice clocks based on optical transitions of neutral atoms, especially for strontium and ytterbium atoms, are the optimal candidates\cite{Cartlidge2018,Riehle2018}. As the accuracy is improved from time to time, the gravity effect on time emerges gradually. The frequency is closely related to the experimental setup where it is located. The uncertainty of the altitude determination is pushed to less than 1 cm\cite{McGrew2018}. Moreover, considering the roadmap for new definition of second, there are five milestones to be satisfied. This also requires the strict measurement of the locations at different gravitational potential\cite{Riehle2018}.

There is a 300-meter-high vertical tunnel built in the ZAIGA-CE-R. It's a good opportunity to verify the LPI experiment with two lattice optical clocks. We plan to put two optical clocks along the vertical tunnel as shown in Fig.~\ref{f1.2}, one is at the summit of the tunnel and the other is at the foot of the tunnel. By comparing the frequency ratios of the two clocks, we can demonstrate the extent of LPI violation at such height.
From the formula above, we can easily get


\begin{equation}
\frac{\Delta f}{f}=(1+\beta)\frac{\Delta U}{c^{2}}=(1+\beta)\frac{g\Delta h}{c^{2}}.
\label{eqn4.3}
\end{equation}

If two optical clocks are used(Fig.~\ref{f1.2}), which keep a stability of 1 part in 10$^{18}$, with $\Delta$\emph{h} =300 m, \emph{g}=9.8 m/s$^{2}$, $\beta$ can be determined at the level of 10$^{-5}$. Moreover, the uncertainty of the height measurement is determined at the level of 10$^{-4}$, and the uncertainty of the gravity acceleration has to be determined at the level of 10$^{-5}$, respectively. So violation of LPI can be tested on the ground 40 years later at the same order of magnitude with GP-A experiment.

\subsection{4.2 ZAIGA-CE-GW prototype}
In the experiment on $^{171}$Yb lattice clocks\cite{McGrew2018}, systematic uncertainty of $1.4 \times 10^{-18}$, and measurement instability of $3.2 \times 10^{-19}$ have been achieved and reproducibility characterized by ten blinded frequency comparisons, yielded a frequency difference of $[-7\pm(5)stat\pm(8)sys] \times 10^{-19}$. This development together with demonstration of systematic uncertainty of Al$^{+}$ optical clock is below 10$^{-18}$ \cite{Brewer1902}, pave the road to 10$^{-21}$ clock precision and accuracy. Proposal of Jun Ye's talk in Urumqi and proposals of others are aiming toward this 10$^{-21}$ clock precision and accuracy. With this clock precision/accuracy, space missions with clocks will are good choice for GW detection in the middle-frequency and low-frequency bands. A space-qualified clock with accuracy of 10$^{-19}$ could already make a good candidate for long arm (longer than 1 AU) GW mission. In the following, we will start with a discussion of radio Doppler tracking of spacecraft\cite{Ni2016,nibook} which could be considered as first GW missions using precision ultra-stable clocks and microwave links, and then go to mission concepts using precision optical clocks and laser links.
In Doppler tracking of S/C, a highly stable master clock on the Earth is used as a reference to control a monochromatic radio wave for transmitting to S/C (uplink). When S/C transponder receives the monochromatic radio wave, it phase-locks the local oscillator with or without a frequency offset and transponds the local oscillator signal back (to the Earth station; downlink) coherently.
The one-way Doppler response \emph{y}(t) is defined as
\begin{equation}
y(t)\equiv\delta\nu/\nu_{0}\equiv(\nu_{1}(t)-\nu_{0})/\nu_{0},
\label{eqn4.4}
\end{equation}
where, $\nu_{0}$ is the frequency of emitted signal and $\nu_{1}$ is the frequency of received signal. Far from the GW sources as it is in the present experimental/observational situations, the plane wave approximation is valid. For weak plane waves propagating in the \emph{z}-direction in GR, we have the following spacetime metric:
\begin{equation}
ds^{2} = dt^{2} – (\delta_{ij} + h_{ij}(ct-z))dx^{i}dx^{j},  |h_{ij}| \ll 1,
\label{eqn4.5}
\end{equation}
where, Latin indices run from 1 to 3 and sum over repeated indices is assumed. Estabrook and Walquist\cite{Estabrook1975,Wahlquist1987} derived the one-way and two-way Doppler responses to plane GWs in weak field approximation Eq.(\ref{eqn4.5}) in the transverse traceless gauge in general relativity. Written in the notation of Armstrong, Estabrook and Tinto\cite{Armstrong1999} the formula for one-way Doppler response on board S/C 2 received from S/C 1 is
\begin{equation}
y(t)=(1 – \vec{k} \cdot \vec{n}) [\Psi(t – (1 + \vec{k} \cdot \vec{n})L) − \Psi(t)],
\label{eqn4.6}
\end{equation}
where, $\vec{k} [= (k^{i}) = (k^{1}, k^{2}, k^{3})]$ is the unit vector in the GW propagation direction, $\vec{n} [= (n^{i}) = (n^{1}, n^{2}, n^{3})]$ the unit vector along the link from spacecraft 1 to spacecraft 2 and \emph{L} is the path length of the Doppler link. The function $\Psi(t)$ is defined as
\begin{equation}
\Psi(t) \equiv n^{i}h_{ij}(t)n^{j} / {2[1 – (\vec{k} \cdot \vec{n})^{2}]}.
\label{eqn4.7}
\end{equation}
With one-way Doppler response known, two-way and multiple way response can easily be written down. As noticed and derived by Tinto and da Silva Alves \cite{Tinto2010}, for GW solutions in any metric theories of gravity of the form Eq.(\ref{eqn4.5}), the Doppler response formula Eq.(\ref{eqn4.6}) and Eq.(\ref{eqn4.7}) are valid also.

Doppler tracking of the Viking S/C (S-band, 2.3 GHz)\cite{Armstrong1979}, the Voyager I S/C (S-band uplink + coherently transponded S-band and X-band (8.4 GHz) downlink) \cite{Hellings1981}, Pioneer 10 (S band)\cite{Anderson1984}, and Pioneer 11 (S band)\cite{Armstrong1987} have been used for GW measurement and have given constraints on GW background in the low-frequency band.

The most recent measurements came from the Cassini spacecraft Doppler tracking (CSDT). Armstrong, Iess, Tortora, and Bertotti\cite{Armstrong2003} used the Cassini multilink radio system during 2001-2002 solar opposition to derive improved observational limits on an isotropic background of low-frequency GWs. The Cassini multilink radio system consists of a sophisticated multilink radio system that simultaneously receives two uplink signals at frequencies of X and Ka bands and transmits three downlink signals with X-band coherent with the X-band uplink, Ka-band coherent with the X-band uplink, and Ka-band coherent with the Ka-band uplink. X band is a standard deep space communication frequency band about 8.4 GHz; Ka band is another deep space communication frequency band about 32 GHz. Armstrong \emph{et al}.\cite{Armstrong2003} used the Cassini multilink radio system with higher frequencies and an advanced tropospheric calibration system to remove the effects of leading noises — plasma and tropospheric scintillation to a level below the other noises. The resulting data were used to construct upper limits on the strength of an isotropic GW background in the 1 μHz to 1 mHz band as \cite{Armstrong2003}: $[S_{h}(f)]^{1/2} < 8 \times 10^{-13}$ at several frequencies in the 0.2$\sim$0.7 mHz band; $h_{c}(f) < 2 \times 10^{-15}$ at frequency about 0.3 mHz; $\Omega_{gw}(f) < 0.03$ at frequency 1.2 $\mu$Hz.

The GW sensitivity of spacecraft Doppler tracking could still be improved by 1-2 order of magnitude with a space borne optical clock on board\cite{Tinto2009}.

In the radio tracking of spacecraft, the received frequency of the signals is tracked. Its integral is the phase. In the radio ranging of spacecraft the received phase of the signals is measured. The derivative of the phase is the frequency. For coherent transponding, the phase measured is basically a ranging up to an additive constant to be determined.

\subsubsection{4.2.1 Pulse laser ranging}
Another way to measure the range is using pulse timing. This is what being done in satellite laser ranging and Lunar laser ranging. Since the fractional accuracies of optical clocks have already reached the 10$^{-18}$ level, when space optical clocks reach this level, pulse laser ranging together with drag-free technology will be an important alternative for detection of GWs in the lower part of low frequency band.

The basic principle of spacecraft Doppler tracking, of spacecraft laser ranging, of space laser interferometers, and of pulsar timing arrays (PTAs) for GW detection are similar. In the development of GW detection methods, spacecraft Doppler tracking method and pulse laser ranging method have stimulated significant inspirations. The methods using space laser interferometers and using PTAs are becoming two important methods of detecting GWs.

\subsubsection{4.2.2 Optical-clock-locked laser-link space GW missions}
These will be high-potential GW mission concept after LISA and other first generation laser interferometric GW missions: using ultra-stable optical clocks to lock lasers of 2 (or more) for transponding to each other (one another) for GW detection in an array of drag-free spacecraft with a distance of AU or more, like ASTROD-GW\cite{Ni2013}, Super-ASTROD\cite{Ni2009}, INO\cite{Ebisuzaki2019}, and the clock mission concepts in references\cite{Loeb2015,Vutha2015,Kolkowitz2016}. At present, good candidates of optical clocks on board are Yb and Sr lattice optical clocks, while on the Earth we can use an ensemble of Yb, Sr, lattice clocks plus Al$^{+}$ ion optical clocks as accuracy calibrator.
\subsubsection{4.2.3 ZAIGA-CE-GW prototype}
In Zhaoshan underground ZAIGA facility, we plan to use 1 km tunnel to set up a 1 km arm length prototype GW detector using optical lattice clocks at the two ends of the prototype and using clock-locked-lasers as link for comparison to test various schemes of GW detection.

\section{5. ZAIGA-RM: Rotation measurement}

In general relativity, the stationary field of a rotating body is different from the static field produced by the same nonrotationg mass, which is a gravito-magnetic effect and consists of a space-time drag due to the mass currents. This rotational frame-dragging effect, also known as the Lense-Thirring effect, was predicted in 1918\cite{Lense1918}. This Lense-Thirring frame dragging effect is important in the understanding of astrophysical phenomena and also in finding the matched templates for detecting gravitational waves. The frame dragging effect on the cryogenic quartz gyroscopes on the Gravity Probe B drag-free satellite was successfully measured\cite{Everitt2011} and Lense-Thirring precession of the orbits of LAGEOS satellites was experimentally verified\cite{Ciufolini2012}. Improving sensitivity of gyroscopes is very important to more precisely measure the Lense-Thirring effect. Large-scale ground laser ring gyroscopes with the aim for geophysical applications and for measuring the Lense-Thirring effect, are under development\cite{Stedman2003,Hurst2004,Schreiber2011,Dunn2002,Belfi2017}. Atom gyroscope has a high potential sensitivity\cite{Scully1993,Dowling1998} and dual Raman-type atom interferometers can cancel the common noise. In the frame of general relativity, the phase shift in a dual-atom-interferometer gyroscope is written as

\begin{equation}
\begin{split}
\delta \phi \cong &4k_{eff}\upsilon T^{2}\Omega_{E}[\cos(\theta+\Psi)-2\frac{GM_{E}}{c^{2}R_{E}}\sin\theta\sin\Psi, \\
                  &+ \frac{GI_{E}}{c^{2}R^{3}_{E}}(2\cos\theta\cos\Psi+\cos\theta\cos\Psi)]
\label{eqn5.1}
\end{split}
\end{equation}
where, $\upsilon$ is the longitudinal velocity of atoms, $\Omega_{E}$ is the rotation rate of the Earth as measured in the local reference frame, $\theta$ is the colatitude of the laboratory, $\Psi$ is the angle between the local radial direction and the normal to the plane of the instrument measured in the meridian plane, \emph{G} is the Newtonian gravitational constant, $M_{E}$ is the mass of the Earth, $R_{E}$ is the position of the laboratory with respect to the center of the Earth, and $I_{E}$ is the moment of inertia of the Earth. In the square bracket, the first term corresponds to the classical Sagnac effect caused by the Earth rotation, the second one is produced by the geodetic effect due to the coupling of gravito-electric field with the rotation of the Earth, and the third contribution is produced by the Lense-Thirring effect due to the gravito-magnetic field caused by the angular momentum of the Earth. In Eq.(\ref{eqn5.1}) two latter terms are general-relativistic effect, which are depressed by a factor of the order of magnitude of 10$^{-9}$ with respect to the classical Sagnac effect. To test the the general relativity effect, above three terms in Eq.(\ref{eqn5.1}) should be separated in the experiment. This could be realized by building a multi-axis atom gyroscope, similar to a multi-ring-laser gyroscope developed for measuring the gravito-magnetic effect\cite{Bosi2011,Beverini2014}. Here, developing atom gyroscopes with an ultrahigh sensitivity becomes very important for measuring the Lense-Thirring effect.

\begin{figure}[htpb]
\includegraphics[width=8.0cm]{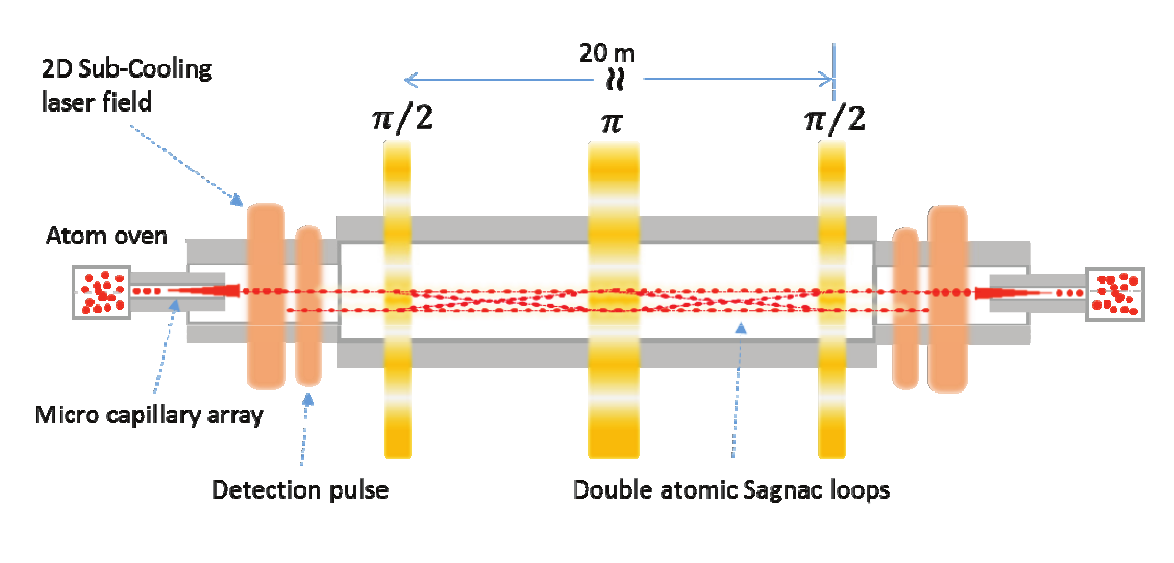}
\caption{Schematic experimental setup of atom gyroscope based on dual atom interferometers for measuring the rotation rate of the earth and even to test the Lense-Thirring effect. \label{f5.1}}
\end{figure}

Benefiting from coherent laser-atom-manipulating techniques, atom gyroscope has been made great progress in the last decades\cite{Gustavson2000,Canuel2006,Stockton2011,Berg2015,Dutta2016,Yao2018}. In the dual-atom-interferometer gyroscope, the rotation resolution ($\delta$$\omega$) is determined by the phase noise of atom interference fringes ($\delta$$\phi$) and the area of interference loop, and it is expressed as
\begin{equation}
\delta\omega = \frac{\delta\phi}{4k_{eff}\upsilon T^{2}}\sqrt{\frac{T_{{\rm c}}}{\tau}},
\label{eqn5.2}
\end{equation}
\begin{equation}
\delta \phi = \frac{1}{C\sqrt{N}},
\label{eqn5.3}
\end{equation}
where, $T_{{\rm c}}$ is the sampling period for each data ($T_{{\rm c}}$$\approx$2$\emph{T}$), $\tau$ is the integrating time for the averaged measurements, \emph{C} is the contrast of interference fringes, and \emph{N} is the number of atoms. As a constituent part of ZAIGA, ZAIGA-RM is a large-scale dual-atom-interferometer gyroscope with a total length of more than 20 meters as shown in Fig.~\ref{f5.1}, it is designed for precisely measuring the rotation rate of the Earth and testing the Lense-Thirring effect, and it is similar to the configuration of Raman-type atom beam gyroscope\cite{Gustavson2000}. Briefly, rubidium atom beams produced by ovens are counter-propagating in a vacuum chamber. After further transversely cooling, atoms are prepared to one ground state. In the interference region, three pairs of counter-propagating Raman lasers are applied to coherently manipulate the atomic wave packets. Thus, two symmetric and overlapped interference loops are built, which allows the rotation dependent phase shift to be observed by detecting the number of atoms in another ground state. A uniform magnetic bias field is applied along the axis of the Raman beams throughout the length of the interference region, and a multi-layer {\rm $\mu$}-metal magnetic shield is used to protect this region from stray magnetic fields. In order to keep more atoms participated in the interference process, the lower atom temperature is necessary as the integrating time increased in our large-scale instrument. Moreover, from Eq.(\ref{eqn5.3}), the total atom number should be as large as possible. Therefore, the design of atom oven with the more atom number and the lower atom temperature is very pivotal in this work.

\begin{figure}[htpb]
\includegraphics[width=8.0cm]{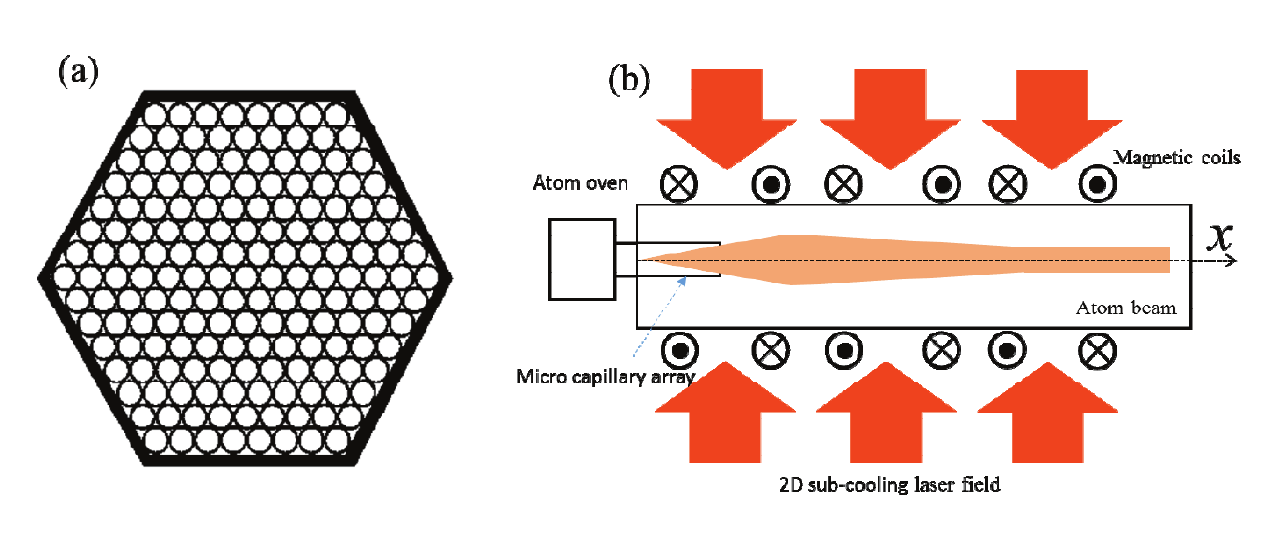}
\caption{Diagram of atom sources including the outlet of the oven and the transverse cooling of atom beam. (a) Cross section of the outlet composed of a micro capillary array. (b) Sub-Doppler cooling of the atom beam using the gradient fields. \label{f5.2}}
\end{figure}

To achieve the persistent atom beams with the high beam intensity and the small divergence angle, outlets of atom ovens are designed with a micro capillary array.\cite{Giordmaine1960} It is composed of 168 capillary tubes arranged in a hexagonal frame, as shown in Fig.~\ref{f5.2}(a). Each capillary with the length of 3 cm has an internal diameter of 150 {\rm $\mu$m} and an external diameter of 200 {\rm $\mu$m}. Based on theoretical estimation\cite{Giordmaine1960,Senaratne2015}, the intensity of atom beam emitting from the outlet of the atom oven is $1.5 \times 10^{14}$ atoms/s, the longitudinal velocity of about 300 m/s, and the temperature of atom oven is 450 K. This design provides the high-intensity atom beams. Due to the small divergence angle, it is convenient for the further transversely sub-Doppler cooling. In addition, the length of the outlet is only 3 cm, the temperature can be conveniently controlled and the outlet would not be blocked. In the large-scale atom gyroscope, the divergence of atom beams becomes more serious due to the longer propagating time of atoms. Thus, the atoms should be further transversely cooled by using the sub-Doppler cooling technique when they emit from the outlet of the atom oven. However, it is difficult for carrying out the transverse sub-Doppler cooling when the longitudinal velocity of atoms is 300 m/s. As shown in Fig.~\ref{f5.2}(b), we design a scheme by using a red-detuning laser and a gradient magnetic field to transversely cool the atoms to near the recoil limit temperature. The spatial magnetic field is distributed as $B_{y} = −ay$, $B_{z} = az$, and $B_{x} = 0$ ($\emph{a}$ is a magnetic field gradient coefficient). When the atom beam is propagating along the $\emph{x}$ direction, they are transversely sub-Doppler cooled. For $^{85}$Rb atoms, the transverse velocity distribution of atom beam by using sub-Doppler technique can be cooled down to 2 cm/s\cite{Balykin2003}.

Compared with the atom gyroscope of a total length of about 2 meters\cite{Gustavson2000}, the ZAIGA-RM with the total length of more than 20 meters has ultrahigh potential sensitivity. Importantly, when the transverse sub-Doppler cooling is applied, the contrast is improved although the total interrogation time is increased. By using Eqs.(\ref{eqn5.2}) and (\ref{eqn5.3}), we evaluate the expected rotation resolution of ZAIGA-RM. Assuming, $\emph{C}$ = 10\%, $\upsilon$ = 300 m/s, $\emph{T}$ = 33 ms, the atom number $\emph{N}$ = $1.0\times10^{13}$ for one sampling period $\emph{T}_{{\rm c}}$ = 66 ms, and $\emph{k}_{{\rm eff}}$ = $1.6\times10^{5}$ cm$^{-1}$, it implies that the rotation resolution ($\delta\omega$) could be achieved to $3.8 \times 10^{-16}$ rad/s when the integrating time $\tau$ = 10000 s. From Eq.(\ref{eqn5.1}), in the location of ZAIGA-RM, based on atom gyroscope analyzed above, if using a multi-axis configuration similar to the multi-ring-laser gyroscope, the Lense-Thirring effect can be measured to less than one percent, the ZAIGA-RM has a promising prospect and it is achievable. The test of the general relativity could also be possible. This work is a primary idea, and the further analysis and design are in progress. In fact, there will be many difficulties in building the ZAIGA-RM. For example, the collimation of the large-scale Raman lasers and symmetry of the large-area interference loops, and also the vibration noise and systematic errors, should be considered.

\section{6. Summary}
Precision measurement plays important role in modern science. Physicists summarize the physical laws through the experimental results obtained by precision measurement and test them under new conditions. The test of WEP is one of the frontiers of gravity differential measurement. At present, the accuracy of the microparticle WEP test is at the level of 10$^{-8}$$\sim$10$^{-9}$. A more accurate test of WEP may lead to the revision of GR or to the support of quantum gravity theory. The WEP test based on the atom interferometer depends on the accuracy of the gravity differential measurement, and thus is proportional to the square of the free evolution time of the atom. An effective measure to extend the free evolution time of atoms on the ground is to increase the baseline of the atom interferometer. The ZAIGA-EP will use a long baseline atom interferometer combined with environmental vibration and noise isolation technology to perform a more accurate WEP test. The ZAIGA-CE-R uses the 10$^{-18}$ level high precision optical frequency standard to accurately measure gravitational redshift, and verify general relativity. Precision measurements of the gravitational field provide information on the temporal and spatial distribution of the Earth's mass. Changes in the Earth's rotation speed and pole shifts result in changes in the distance between the gravitational measurement point and the Earth's axis, the ZAIGA-GM uses high-precision atomic gravimeter, atom gyro, and laser interferometer to accurately measure geophysical parameters such as Earth's gravitational field, Earth's rotation, Earth's strain field, and is used for basic physics research such as gravitational magnetic effects. The ZAIGA-GW is expected to detect the middle-frequency-band gravitational waves and answer the question of whether there are IMBHs. In summary, the construction of the ZAIGA facility can provide a basis for testing basic physical laws, gravitational redshift effects, gravitational magnetic effects, and detecting mid-range gravitational waves.

\begin{acknowledgments}
We acknowledge P. Buoyer and G. M. Tino for their helpful discussion. This work was supported by National Key Research and Development Program of China under Grant Nos. 2016YFA0302002 and 2017YFC0601602, National Natural Science Foundation of China under Grant Nos. 91536221, 91736311, 11574354 and 11204354, and Strategic Priority Research Program of the Chinese Academy of Sciences under grant No. XDB21010100.
\end{acknowledgments}

\end{document}